\documentclass[aps,prl,reprint,showpacs,showkeys,groupedaddress,nofootinbib,superscriptaddress]{revtex4-1}

\usepackage{hyperref}
\usepackage[T1]{fontenc}
\usepackage[ansinew]{inputenc}
\usepackage{amsmath}
\usepackage{setspace}
\usepackage{psfrag}
\usepackage{subfigure}
\usepackage{mathtools}
\usepackage{amssymb,lineno,amsfonts}
\usepackage{verbatim}
\usepackage{color}
\usepackage{textcomp}
\usepackage[toc,page]{appendix}
\usepackage{graphicx}
\usepackage{bm}
\usepackage{dcolumn}
\usepackage{gensymb}
\usepackage{ifpdf}
\usepackage{bbm}
\usepackage{mathrsfs}
\usepackage{upgreek}
\usepackage{epstopdf}
\usepackage{esvect}
\usepackage[usenames,dvipsnames]{xcolor}
\definecolor{med-blue}{RGB}{25,25,112}
\hypersetup{colorlinks, linkcolor={blue},citecolor={blue}, urlcolor={blue}}
\usepackage[english]{babel}
\usepackage[autostyle, english = american]{csquotes}
\MakeOuterQuote{"}
\newcommand{\be}{\begin{equation}}
\newcommand{\ee}{\end{equation}}
\newcommand{\ben}{\begin{equation*}}
\newcommand{\een}{\end{equation*}}
\newcommand{\bea}{\begin{eqnarray}}
\newcommand{\eea}{\end{eqnarray}}
\newcommand{\bean}{\begin{eqnarray*}}
\newcommand{\eean}{\end{eqnarray*}}
\newcommand{\ti}{\textit}
\newcommand{\tb}{\textbf}
\newcommand{\tu}{\underline}
\newcommand{\na}{\mbox{\boldmath$\nabla$}}
\begin{document}

\title{Quantum Potential induced UV-IR coupling in Analogue Hawking radiation: From Bose-Einstein Condensates to canonical acoustic black holes}

\author{Supratik Sarkar}
\thanks{Electronic address: supratiks@students.iiserpune.ac.in}
\author{A. Bhattacharyay}
\thanks{Electronic address: a.bhattacharyay@iiserpune.ac.in}

\affiliation{Indian Institute of Science Education and Research, Pune 411008, India}

\date{\today}

\begin{abstract}
Arising out of a Non-local non-relativistic BEC, we present an Analogue gravity model upto $\mathcal{O}(\xi^{2})$ accuracy in the presence of the quantum potential term for a canonical acoustic BH in $(3+1)$-d spacetime where the series solution of the free minimally coupled KG equation for the large length scale massive scalar modes is derived. We systematically address the issues of the presence of the quantum potential term being the root cause of a UV-IR coupling between short wavelength \ti{primary} modes which are supposedly Hawking radiated through the \textcolor{black}{sonic} horizon and the large wavelength \ti{secondary} modes. In the quantum gravity experiments of analogue Hawking radiation in the laboratory, this UV-IR coupling is inevitable and one can not get rid of these large wavelength excitations which would grow over space by gaining energy from the short wavelength Hawking radiated modes. We identify the characteristic feature in the growth rate(s) that would distinguish these primary and secondary modes. 
\end{abstract}


\keywords{analogue gravity, BEC, quantum potential, massive Klein-Gordon, acoustic black holes, Ultraviolet - Infrared coupling, Hawking radiation}

\maketitle

\section{1. Introduction}
It is well-known that from an experimentally realizable condensed matter model, through some rigorous mathematical framework, Gravity comes out as an \ti{emergent phenomenon} as seen by the sonic excitations. Unruh's seminal work \cite{Unruh1981} practically opened up this field of research which has been extensively pursued over a last couple of decades and having a host of theoretical proposals around \cite{Jacobson1991,Barcelo2011}. 
\par 
Among many other "analogue models", from the standpoint of classical physics, the passing of sound waves as acoustic disturbances through a moving Newtonian fluid is the simplest and cleanest example of the condensed matter analog for the light waves in a curved spacetime \cite{Unruh1995,Visser1993,Visser1998}. The idea is if the fluid flow ever becomes supersonic then, in that trapped region, the sound waves would never be able to fight their way back upstream and this surface of no-return in the fluid medium clearly bears the analogy of gravitational event horizon. \textcolor{black}{Although this terminology of \ti{acoustic analogue of event horizon} (i.e. sonic horizon) doesn't qualify by the stricter definition of \ti{event horizon} in General Relativity, nevertheless, this is actually a \ti{Killing horizon} from which an analogue Hawking radiation can be expected.} This implies the very existence of a "dumb hole" or, in other words, \ti{acoustic black hole} \cite{Unruh2008,Visser1999}.
\par
One can probe various aspects of curved spacetime Quantum Field Theory (QFT) via these analogue models due to the amenability of accurate experimental control and observational verification - a quantum system characterized by very cold temperature ($\sim$100 nK), low speed of sound and high degree of quantum coherence offers the best test field \cite{Garay2000,Garay2001}. Bose-Einstein Condensate (BEC), which is a superfluid quantum phase of matter, happens to be one of the most prominent candidates of all such ultra cold systems \cite{Visser1998,Barcelo2003} to examine and investigate some crucial features of \ti{Emergent Gravity}. It does create analogue gravitational scenarios at nK temperatures within the laboratory setup and readily provides with an experimental window to capture some key aspects of Hawking radiation\footnote{In order for a gravitational black hole (with mass, say, $M_{\text{\tiny{BH}}}$) to be observed to emit Hawking radiation, it must have a temperature (say, $T_{\text{\tiny{H}}}$) greater than that of the present-day Cosmic Microwave Background (CMB) radiation (say, $T_{\text{\tiny{CMB}}}$). $T_{\text{\tiny{CMB}}}=2.725\hspace{0.03cm}\text{K}$, while $T_{\text{\tiny{H}}}=\frac{\hbar c^{3}}{8\pi G M_{\text{\tiny{BH}}}k_{\text{B}}} \approx 6.169 \times 10^{-8}\hspace{0.03cm}\text{K} \times \frac{M_{\text{\tiny{sun}}}}{M_{\text{\tiny{BH}}}}$ in SI units; and hence the direct detection of the gravitational Hawking radiation for a Schwarzschild black hole with a mass equivalent to at least the solar mass becomes far below beyond the limit of the current observational techniques. So its extremely difficult to verify Hawking radiation in nature and the reason being $T_{\text{\tiny{H}}}$ is seven orders of magnitude smaller than $T_{\text{\tiny{CMB}}}$.} which is one of the cornerstone results \cite{Birrell1982} of curved spacetime QFT. 
\par 
This fact basically led to the increasing interest in using BEC \footnote{Apart from BEC, some other condensed matter systems, such as superfluid helium \cite{VOLOVIK1996,Jacobson1998}, superconductors \cite{Ge2010,Ge2012}, polariton superfluid \cite{Gerace2012} and degenerate Fermi gas \cite{Giovanazzi2005}, have also been used as the tools to probe Emergent Gravity} as a platform to observe \ti{analogue} Hawking radiation \cite{Fedichev2003,Zapata2011,Lahav2010} as a thermal bath of phonons with the temperature proportional to the \ti{surface gravity} \cite{Barcelo2011,Visser1998}. In this context, Parentani and co-workers have already proposed some novel ideas based on density correlations, studying the hydrodynamics over several length scales and even surface-gravity-independent-temperature etc., \cite{Macher2009,Finazzi2011a,Finazzi2011b,Finazzi2012}. In order to experimentally detect analogue Hawking radiation, a stack of recent works are obviously worth mentioning here \cite{Belgiorno2010,Schutzhold2011,Marino2008,Marino2009,Fouxon2010,Solnyshkov2011,Liberati2012,Finazzi2012b,Steinhauer2014}.
\par
In a BEC, the small amplitude collective excitations \cite{Pitaevskii2003} of the uniform density moving phase (to be precise, the first order phase-fluctuation-field) obeys the quantum hydrodynamics which, ignoring the "Quantum Potential" term (refer to Section-4.2.1 of \cite{Barcelo2011}), can be cast into the d'Alembertian equation of motion of a massless minimally coupled free scalar field on a $(3+1)$-d Lorentzian manifold with an effective metric to be regarded as the acoustic metric of the curved background\footnote{Refer to eqn.(254) of \cite{Barcelo2011}.}. 
\par 
Some worthwhile efforts have been taken in order to regularize the dynamics taking the quantum potential term into account \cite{Fleurov2012}. In 2005, Visser \ti{et al.} had shown the emergence of a massive Klein-Gordon (KG) equation considering a two-component-BEC where a laser induced transition between the two components was exploited \cite{Visser2005}. Liberati \ti{et al.} proposed a weak $U(1)$ symmetry breaking of the analogue BEC model by the introduction of an extra quadratic term in the Hamiltonian to make the scalar field massive \cite{Liberati2009}. Considering the flow in a Laval nozzle, Cuyubamba has shown the emergence of a massive scalar field in the context of analogue gravity arguing for the possibility of observation of quasi-normal ringing of the massive scalar field within the laboratory setup \cite{Cuyubamba2013}. \textcolor{black}{In the context of Trans-Planckian backaction issues on low-energy predictions in Analogue Gravity, Fischer \ti{et al.} has recently done a remarkable work \cite{Fischer2017}.}
\par 
\textcolor{black}{It is well known that, in order to obtain an analogue gravity model from a non-relativistic BEC, one necessarily needs to get rid of the terms coming from the linearisation of the quantum potential to find the effective acoustic metric on the basis of ``hydrodynamical'' approximation where the contributions coming from the small length scale (wavelengths shorter than the healing length $\xi$ of the system) regime are neglected. However if one goes beyond this hydrodynamic regime to access the high frequency modes, the contribution of the quantum potential can not be neglected and the acoustic description can still be achieved through \ti{Eikonal approximation} - the Lorentz breaking in BEC models \cite{Barcelo2011}. 
 The main advantage of this approximation is that the ``operator'' $\hat{D}_{2}$ (refer to Eq.\eqref{eq:aa09} later) can effectively be replaced by just the ``function'' and consequently the entries of the acoustic metric become explicitly momentum dependent numbers, but not operators. In this case, it gives rise to a modified dispersion relation for the quasiparticles, refer to eqn.(271) of \cite{Barcelo2011}. In this regard of the significance of the Eikonal approximation, the points $4$ and $5$ as mentioned by Barcel\'o \ti{et al.} on p.64 of \cite{Barcelo2011} are of great importance.
}
\par
In our previous paper \cite{Sarkar2016}, we have looked into the effect of the Lorentz-breaking quantum potential term in a different way. This is a term of immense importance in the context of analogue gravity because this gives rise to the dispersion relation which is used to present an alternative scenario of the analogue Hawking radiation bypassing the trans-Planckian problem \cite{Macher2009,Finazzi2011a,Finazzi2011b,Finazzi2012,Corley1996,Corley1998}. But the presence of this quantum potential term in the dynamics is somewhat analogous to that of a diffusion term which should spread the small scale modes into the large scale ones. In our previous work, we had guessed the existence of this coupling between the small and large scale dynamics and had captured this picture through the massive large wave length excitations as the amplitude modes over the usual small scale excitations. This whole work was presented on flat spacetime for the sake of simplicity and in order to introduce the idea.
\par 
In the present paper, we analyze in details, the effect and consequence of the presence of the quantum potential term in the context of spreading out of the small scale excitations into the large scale ones on the curved spacetime of a canonical (a model first proposed by Visser \cite{Visser1998} in 1998) acoustic black hole. This is an important analysis in its own right, because, the Ultra-Violet (UV) to Infra-Red (IR) coupling is inevitable in these types of systems. This, in turn, results in the presence of instabilities to the short wavelength (UV correspondence) modes which are predominant in the Hawking spectrum as seen by a free-falling observer in a local Minkowski spacetime. This is because of the fact that, at very large curvatures, the local Minkowski flat space can only account for the short wavelength (UV correspondence) modes. The large wavelength (IR correspondence) modes which would be subsequently generated out of the UV-modes are characterized by a mass term  solely dependent on the small scales. This fact does manifest the energetic dependence of these IR-modes on the UV ones and, at some sufficiently large time, there would be a transfer of energy from the UV-modes to the IR ones and this can completely mask the Hawking signal of the Hawking radiated modes. A detailed analysis of the UV-IR coupled dynamics is quite essential in that respect. Very recently, Vieira \ti{et al.} have discussed the analogue Hawking radiation of the massless scalar particles and the features of the Hawking spectrum associated in the sapcetime of rotating and canonical acoustic black holes \cite{Vieira2016}.
\par 
Here we have obtained the series solution to the free minimally coupled "massive" KG equation on a $(3+1)$-d canonical curved background\footnote{Refer to the line element given by eqn.(55) of ref.\cite{Visser1998}}. And we have followed the method shown by Elizalde \cite{Elizalde1988} in 1988. The acoustic situation although being similar to the Schwarzschild geometry, however, the metric (i.e. acoustic metric) itself is quite different from the standard Schwarzschild metric. As a result, our present study here is quite different in some details from the one done by Elizalde on the Scwarzschild background.
\par 
Using the series solution truncated to the desired accuracy, we show that the IR-modes, having grown from the UV-modes (supposedly Hawking radiated), would have a \textcolor{black}{dominant} \ti{power law growth over space} which is characterized by the quartic power of the UV-frequency of the original Hawking radiated modes. This has the clear indication that there remains the information of relative abundance of the Hawking radiated quanta within the growth of the IR-modes generated by the UV Hawking radiated modes. 
\par 
In view of the inevitability of the appearance of the UV-IR coupling through the presence of the quantum potential term in these analogue gravity models, the present exercise of ours does provide a better look at the prevailing situation in such systems.
\par 
This paper is organized in the following manner - in \ti{Section 2}, we start from the stepping stone which is the nonlocal Gross-Pitaevskii (GP) model for a BEC with interaction. By adopting the Madelung ansatz and adding the first order fluctuations to the density and phase of the single-particle BEC state, we develop a setting from the condensed matter standpoint.
\par  
In \ti{Section 3}, we give the structure of the full model as proposed by us, derive the general acoustic metric of a $(3+1)$-d curved background and form the covariant "massive" KG equation for a free scalar field.
\par 
In \ti{Section 4}, we consider the geometry of a canonical acoustic black hole where the fluid flow is assumed to be incompressible and spherically symmetric and we derive the exact solution to the massive KG equation on this background, taking into account the radial part specifically. 
\par 
In \ti{Section 5}, we conclude the paper with a detailed discussion of our results.

\section{2. The Setting}

\textcolor{black}{To study interacting non-uniform Bose gases at very low temperature, one uses the Bogoliubov prescription for the field operator $\hat{\psi}(t,\tb{r})$ that obeys the well-known commutation relations $\left[ \hat{\psi}(t,\tb{r}), \hspace{0.05cm}\hat{\psi}^{\dagger}(t,\tb{r}^{\prime})\right]=\delta(\tb{r}-\tb{r}^{\prime})$ and $\left[ \hat{\psi}(t,\tb{r}), \hspace{0.05cm}\hat{\psi}(t,\tb{r}^{\prime})\right]=0$. In the Heisenberg picture, the dynamics of this Bose-field-operator $\hat{\psi}(t,\tb{r})$ is given by the exact equation\footnote{See eqn.(5.1) of \cite{Pitaevskii2003}.}}
\begin{widetext}
\textcolor{black}{
\be \label{eq:rev1}
 i\hbar \hspace{0.05cm}\partial_{t} \hspace{0.05cm} \hat{\psi}(t,\tb{r}) = \left[ \hat{\psi}(t,\tb{r}), \hspace{0.05cm}\hat{\mathcal{H}}\right] = \left( -\frac{\hbar^{2}}{2m}\na^{2} + V_{\text{ext}}(t,\tb{r}) + \int \hat{\psi}^{\dagger}(t,\tb{r}^{\prime}) \hspace{0.04cm}V(\tb{r}^{\prime}-\tb{r}) \hspace{0.04cm}\hat{\psi}(t,\tb{r}^{\prime}) \hspace{0.04cm}d\tb{r}^{\prime} \right) \hspace{0.05cm}\hat{\psi}(t,\tb{r})
 \hspace{0.1cm}, 
\ee
}
\end{widetext}
\textcolor{black}{where the many-body Hamiltonian $\hat{\mathcal{H}}$ of the full interacting Bose-system, as inserted above, is given by }
\textcolor{black}{
\bea \nonumber
&\hat{\mathcal{H}} = \int \left(\frac{\hbar^{2}}{2m}\na \hat{\psi}^{\dagger}(t,\tb{r}) \hspace{0.04cm}\na \hat{\psi}(t,\tb{r}) \right) \hspace{0.05cm}d\tb{r} \\ \nonumber 
&\hspace{0.05cm}+  \int \hat{\psi}^{\dagger}(t,\tb{r}) \hspace{0.04cm} V_{\text{ext}}(t,\tb{r})\hspace{0.04cm} \hat{\psi}(t,\tb{r})  \hspace{0.06cm}d\tb{r} \\ \label{eq:rev2} 
&\hspace{0.05cm}+ \frac{1}{2}\int \hat{\psi}^{\dagger}(t,\tb{r}) \hspace{0.04cm}\hat{\psi}^{\dagger}(t,\tb{r}^{\prime}) \hspace{0.04cm}V(\tb{r}^{\prime}-\tb{r}) \hspace{0.04cm}\hat{\psi}(t,\tb{r}) \hspace{0.04cm}\hat{\psi}(t,\tb{r}^{\prime}) \hspace{0.04cm}d\tb{r}^{\prime} d\tb{r} \hspace{0.05cm}, \hspace{0.7cm}
\eea
with $\hbar=h/2\pi$, $h$ being the Planck constant, $m$ being the mass of a single boson. Obviously, $V_{\text{ext}}(t,\tb{r})$ is an external (trapping) potential and $V(\tb{r}^{\prime}-\tb{r})$ is the interaction potential.
}
\par
\textcolor{black}{
Now to the lowest order Born approximation and at very low temperatures, one gets the license to replace the quantum field operator $\hat{\psi}(t,\tb{r})$ by the classical wave function $\psi(t,\tb{r})$ of the condensate due to the macroscopic occupation of a large number of atoms in a single quantum state (BEC ground state). The mean field approximation is 
\be \label{eq:rev3}
\hat{\psi}(t,\tb{r}) \hspace{0.1cm}\rightarrow \hspace{0.2cm}<\hat{\psi}>\hspace{0.05cm}=\psi(t,\tb{r}) \hspace{0.5cm},
\ee
by which one sort of neglects the noncommutativity of the field operators $\hat{\psi}(t,\tb{r})$ as defined above. This mean-field approximation has its implication from a physical point of view as well - since $\hat{\psi}(t,\tb{r})$ or $\hat{\psi}^{\dagger}(t,\tb{r})$
act as `annihilation' or `creation' operator(s) respectively to annihilate or create a particle at $(t, \tb{r})$, now if a particle is subtracted from or added to the condensate, it does not really change the physical properties of the whole system which is actually governed by the order parameter $\psi(t,\tb{r})$. Moreover, this switching from $\hat{\psi}(t,\tb{r})$ to the classical mean-field $\psi(t,\tb{r})$ is accurate enough when one does not consider the realistic potential\footnote{See the argument on p.39 of \cite{Pitaevskii2003}.} but replaces $V(\tb{r}^{\prime}-\tb{r})$ by some effective soft potential $V_{\text{eff}}(\tb{r}^{\prime}-\tb{r})$.
}
\par
The minimal GP model for the \ti{nonlocal} \cite{Sarkar2014} \ti{s}-wave scattering in a non-uniform BEC is characterised by the following equation,
\bea\nonumber 
& i\hbar \hspace{0.05cm}\partial_{t} \hspace{0.05cm} \psi(t,\tb{r}) = \left( -\frac{\hbar^{2}}{2m}\na^{2} + V_{\text{ext}}(t,\tb{r}) + \mathsf{g}|\psi(t,\tb{r})|^{2}\right) \psi(t,\tb{r})  \\ \label{eq:aa01}
& \hspace{3cm}+ \kappa a^{2}\mathsf{g}\psi(t,\tb{r})\na^{2}|\psi(t,\tb{r})|^{2} \hspace{0.1cm},
\eea
where $\psi(t,\tb{r})$ is the condensate wave function which plays a role of the order parameter of the system and thus $|\psi(t,\tb{r})|^{2}=n(t,\tb{r})$ is the density of the condensate, $\mathsf{g}=4\pi\hbar^{2} a/m$ parameterises the strength of the $s$-wave scattering (between different bosons in the gas) considered at the lowest order Born approximation with $a$ being the $s$-wave scattering length and $\kappa$ is some numerical pre-factor of the nonlocal correction term corresponding to the specific coordinate system under consideration \textcolor{black}{(for instance, $\kappa=1/2$ for a 3-d Cartesian system \cite{Sarkar2016})}.
\par
\textcolor{black}{The interaction term $\mathsf{g}|\psi(t,\tb{r})|^{2}\psi(t,\tb{r})$ in the local\footnote{See eqn.(5.2) of \cite{Pitaevskii2003}.} GP equation comes by considering a $\delta$-function approximation to the interaction potential $V(\tb{r}^{\prime}-\tb{r})$ in the $s$-wave interaction picture $\int \psi^{*}(t,\tb{r}^{\prime}) V(\tb{r}^{\prime}-\tb{r}) \psi(t,\tb{r}^{\prime}) d\tb{r}^{\prime}$ in a non-uniform BEC; i.e., $V(\tb{r}^{\prime}-\tb{r})\equiv V_{\text{eff}}(\tb{r}^{\prime}-\tb{r})=\mathsf{g}\delta(\tb{r}^{\prime}-\tb{r})$ is substituted. This approximation is valid under the consideration $|a|<<n^{-\frac{1}{3}}$ which is the condition of diluteness, i.e., the $s$-wave scattering length (that characterises all the effects of Boson-Boson interaction on the physical properties of the gas) is much smaller than the average interparticle separation. Now in a BEC, one can tune the $s$-wave scattering length by Feshbach resonance (see the arguments later where $\epsilon$ would be introduced in Eq.\eqref{eq:aa11}) and this actually opens up the possibility of going away from the diluteness limit of $|a|<<n^{-\frac{1}{3}}$. Keeping this in mind, a correction term (i.e. the last term in Eq.\eqref{eq:aa01}) can be derived to bring in the effects of nonlocality of the interactions. A Taylor expansion of $\psi(t,\tb{r}^{\prime})$ about $\tb{r}^{\prime}=\tb{r}$ in the interaction term $\int \psi^{*}(t,\tb{r}^{\prime}) V(\tb{r}^{\prime}-\tb{r}) \psi(t,\tb{r}^{\prime}) d\tb{r}^{\prime}$ will give rise to the minimal correction at the second order since the first order correction vanishes due to spherical symmetry. 
}
\par 
\textcolor{black}{We have given a detailed description of the derivation of the above mentioned Gross-Pitaevskii equation with a non-local correction term in our previous paper \cite{Sarkar2016}.} In Eq.\eqref{eq:aa01}, the last term on r.h.s represents the minimal nonlocal "correction" to the standard GP equation with contact interactions. If the width of the interaction potential be of the order of a microscopic length scale, say $\varsigma$, then the minimal correction term due to the nonlocality arises out to be $\propto \hspace{0.05cm} \varsigma^{2}$. For the sake of simplicity, here we have taken the interaction-width to be of the order of the $s$-wave scattering length, i.e. $\varsigma \hspace{0.05cm}\simeq \hspace{0.05cm}a$, which won't affect any of our results qualitatively.

\par
On considering a general single particle state of the BEC as 
\be \label{eq:aa02}
\psi(t,\tb{r})=\sqrt{n(t,\tb{r})}\hspace{0.05cm}e^{i\vartheta(t,\tb{r})/\hbar}
\ee 
by adopting the Madelung ansatz, Eq.\eqref{eq:aa01} gives rise to a set of coupled equations\footnote{Here, Euler equation (Eq.\eqref{eq:aa04}) is written in Hamilton-Jacobi form for convenience. One may refer to eqn.(5.15) of \cite{Pitaevskii2003}.};
\begin{widetext}
\begin{align} \label{eq:aa03}
&\text{\small{Continuity equation:}} \hspace{3cm} \partial_{t}\hspace{0.05cm}n(t,\tb{r})+\frac{1}{m}\na.\Big(n(t,\tb{r})\na \vartheta(t,\tb{r}) \Big)=0, \\ \nonumber \\ \label{eq:aa04}
&\text{\small{Euler equation:}} \hspace{1cm} \partial_{t}\hspace{0.05cm}\vartheta(t,\tb{r})+\left(\frac{\left[\na \vartheta(t,\tb{r})\right]^{2}}{2m}+V_{\text{ext}}(t,\tb{r})+\mathsf{g}n(t,\tb{r})-\frac{\hbar^{2}}{2m}\frac{\na^{2}\sqrt{n(t,\tb{r})}}{\sqrt{n(t,\tb{r})}} + \kappa a^{2}\mathsf{g}\hspace{0.05cm}\na^{2}n(t,\tb{r}) \right)=0.
\end{align}
\end{widetext}
\textcolor{black}{Its quite evident that due to the presence of the nonlocal correction term in Eq.\eqref{eq:aa01}, the ``modified'' form of the Quantum Potential as obtained from the above Eq.\eqref{eq:aa04} is given by
\be \label{eq:new05}
\mathsf{V}_{\text{quantum}} = -\frac{\hbar^{2}}{2m}\frac{\na^{2}\sqrt{n(t,\tb{r})}}{\sqrt{n(t,\tb{r})}} + \kappa a^{2}\mathsf{g}\hspace{0.05cm}\na^{2}n(t,\tb{r}).
\ee
}
At this stage, we purposely define two independent scales in spherical polar coordinates as
\be \label{eq:aa05}
x \equiv x^{\mu}=(t, r, \theta, \phi) \hspace{0.2cm} \text{and} \hspace{0.2cm} X \equiv X^{\bar{\mu}}=(\mathcal{T}, R, \Theta, \Phi);
\ee
viz. the small scale and large scale respectively. Here $\mu$, $\nu$,... etc. along with $\bar{\mu}$, $\bar{\nu}$,... etc. are the two different sets of free/dummy indices which separately run over the small and large scales respectively. Later in this section, we will introduce the corresponding spacetime derivatives as well.
\par
Now we consider the fluctuations\footnote{These linearised fluctuations in the dynamical quantities are more generally referred to as \ti{acoustic disturbances}. To be precise and according to the convention, the low-frequency large-wavelength disturbances are called \ti{wind gusts}, while the high-frequency short-wavelength disturbances being described as acoustic disturbances. Refer to \cite{Visser1998}-p.1770.} to the density (i.e. $n(t, \tb{r})$) and phase (i.e. $\vartheta(t, \tb{r})$) of the BEC state in the following manner,
\begin{align} \label{eq:aa06}
n \rightarrow n_{0} + n_{1}(X, x) \hspace{0.3cm},\hspace{0.3cm}  & \vartheta \rightarrow \vartheta_{0}(x) + \left\lbrace \vartheta_{2}(X)\vartheta_{1}(x) \right\rbrace ;
\end{align}
i.e., $n_{0}$ and $\vartheta_{0}(x)$ are basically the classical mean-field density and phase respectively, \textcolor{black}{such that $<n>=n_{0}$ and $<\vartheta>=\vartheta_{0}(x)$. These are obviously the macroscopic descriptions of the condensate within the classical regime.} On the other hand, $n_{1}(X,x)$ is the first order density-fluctuation and $\left\lbrace \vartheta_{2}(X)\vartheta_{1}(x) \right\rbrace $ is the first order phase-fluctuation (accompanying amplitude modulations). In other words, these fluctuations are the quantum fields in nature and can be described as "quantum acoustic representation" \footnote{Refer to eqn.(242) of \cite{Barcelo2011}.}.
\par
Inserting Eq.\eqref{eq:aa06} back into Eq.\eqref{eq:aa03} and Eq.\eqref{eq:aa04}, one comes up with the linearised dynamics. \textcolor{black}{It is to be carefully noted here that, at this stage, the usual partial derivatives do involve independent multiple scales which are mixed and clubbed together as of now. Since we are not operating these derivatives onto the density and/or phase fields at the moment, we may opt to refrain ourselves from the necessary modification of the derivative operator(s) for notational convenience. But to keep everything notationally pellucid, these usual partial derivative operators are just being ``renamed'' by the same with a tilde over each one for now. However the modification of the derivative operator(s) is explicitly performed later in Eq.\eqref{eq:aa18}.}
\par
Considering $V_{\text{ext}}(t, \tb{r})=0$ and $n_{0}$ as just a constant for simplicity, the linearised dynamics is given by Eq.\eqref{eq:aa07} (obtained from the continuity equation) coupled with Eq.\eqref{eq:aa08} (obtained from the Euler equation) as the following,
\begin{small}
\begin{align} \label{eq:aa07}
& \tilde{\partial}_{t} \hspace{0.05cm}n_{1}(X, x)+\frac{1}{m}\tilde{\na} . \bigg(n_{1}(X, x)\tilde{\na} \vartheta_{0}(x) + n_{0}\tilde{\na} \{\vartheta_{2}(X)\vartheta_{1}(x) \}\bigg)=0,\\ \nonumber
 & \mbox{and} \\ \nonumber
& \tilde{\partial}_{t} \hspace{0.05cm}\{\vartheta_{2}(X)\vartheta_{1}(x)\} + \frac{1}{m}\tilde{\na} \vartheta_{0}(x) . \tilde{\na}\{\vartheta_{2}(X)\vartheta_{1}(x)\} + \mathsf{g}n_{1}(X, x) \\ \label{eq:aa08}
& \hspace{5cm}- \frac{\hbar^{2}}{2m}\hat{D}_{2} \hspace{0.08cm} n_{1}(X, x)=0,
\end{align}
\end{small}
where 
\be \label{eq:aa09}
\hat{D}_{2} \equiv \frac{2m\mathsf{g}}{\hbar^{2}}\left( \frac{\hbar^{2}}{4m \mathsf{g} n_{0}}- \kappa a^{2}\right ) \tilde{\na}^{2} = \frac{2m\mathsf{g}}{\hbar^{2}}\xi^{2}\tilde{\na}^{2}
\ee
represents a second-order differential operator obtained by linearising \textcolor{black}{$\mathsf{V}_{\text{quantum}}$ from Eq.\eqref{eq:new05}} and $\xi$ is the ``modified'' healing length corresponding to our proposed \ti{nonlocal} model, given by,
\be \label{eq:aa10}
\xi = \left(\frac{\hbar^{2}}{4 m \mathsf{g} n_{0}} - \kappa a^{2} \right)^{1/2} = \xi_{0}\hspace{0.05cm}\epsilon\hspace{0.2cm}; \hspace{0.2cm} \xi_{0}=\frac{\hbar}{\sqrt{2m\mathsf{g}n_{0}}}.
\ee
\\
\\
This $\xi_{0}$ is nothing but the healing length\footnote{Refer to eqn.(5.20) of \cite{Pitaevskii2003}.} corresponding to the usual \ti{local} GP model and $\epsilon$ is a parameter set to be a very small quantity, 
\be \label{eq:aa11}
\epsilon = \left(\frac{1}{2} -  8\pi \kappa a^{3}n_{0} \right)^{1/2}.
\ee 
\textcolor{black}{The $s$-wave scattering length $a$ can practically be increased from $-\infty$ to $\infty$ near a \ti{Feshbach resonance} as experimentally verified by Cornish \ti{et al.} in 2000 \cite{Cornish2000}.} Evidently, the tuning of $a$ will keep increasing or decreasing the value of the parameter $\epsilon $ as per requirement \cite{Sarkar2014,Barcelo2003a}. \textcolor{black}{In fact, by increasing $a$ through Feshbach resonance, the value of $8\pi \kappa a^{3}n_{0}$ can be made as close to $\frac{1}{2}$ as possible and hence, naturally, $\epsilon$ \textcolor{black}{can be experimentally set} to be a very small quantity via Eq.\eqref{eq:aa11}.}
\par 
From Eq.\eqref{eq:aa08}, \hspace{0.05cm} $n_{1}(X, x)$ can be obtained as the following,
\bea \nonumber
& n_{1}(X, x) \hspace{7cm}\\ \nonumber
& \hspace{0.5cm}= - \hat{\mathcal{A}}
\bigg(\tilde{\partial}_{t} \hspace{0.05cm}\{\vartheta_{2}(X)\vartheta_{1}(x)\} + \frac{1}{m}\tilde{\na} \vartheta_{0}(x) . \tilde{\na}\{\vartheta_{2}(X)\vartheta_{1}(x)\} \bigg), \\ \nonumber 
 & \text{where,}\hspace{0.5cm} \hat{\mathcal{A}}=\Big(\mathsf{g} - \frac{\hbar^{2}}{2m}\hat{D}_{2}\Big)^{-1} \simeq \hspace{0.3cm}\mathsf{g}^{-1}\Big(1+\xi^{2} \tilde{\na}^{2}\Big). \\ \label{eq:aa12}
\eea
Here $\epsilon$ being a very small quantity (as given by Eq.\eqref{eq:aa11}) is exactly what \textcolor{black}{would} allow us to take a binomial approximation above. This expression of $n_{1}(X, x)$ is again substituted back into the Eq.\eqref{eq:aa07} to get a second order partial differential equation in terms of the phase fluctuations, i.e.,
\begin{widetext}
\bea \nonumber
&\tilde{\partial}_{t} \hspace{0.05cm}\bigg[-\hat{\mathcal{A}} \hspace{0.1cm}
\Big(\tilde{\partial}_{t} \hspace{0.05cm}\{\vartheta_{2}(X)\vartheta_{1}(x)\} + \frac{1}{m}\tilde{\na} \vartheta_{0}(x) . \tilde{\na}\{\vartheta_{2}(X)\vartheta_{1}(x)\} \Big) \bigg]
\\ \label{eq:aa13}
&+\frac{1}{m}\tilde{\na} . \bigg(\bigg[-\hat{\mathcal{A}} \hspace{0.1cm}
\Big(\tilde{\partial}_{t} \hspace{0.05cm}\{\vartheta_{2}(X)\vartheta_{1}(x)\} + \frac{1}{m}\tilde{\na} \vartheta_{0}(x) . \tilde{\na}\{\vartheta_{2}(X)\vartheta_{1}(x)\} \Big) \bigg]\tilde{\na} \vartheta_{0}(x) + n_{0}\tilde{\na} \{\vartheta_{2}(X)\vartheta_{1}(x) \}\bigg)=0.  
\eea
\end{widetext}

In the present context of passing a sonic disturbance through a barotropic inviscid fluid\footnote{For a detailed justification, refer to \cite{Barcelo2011}-p.9.}, the background fluid flow $\tb{v}(x)$ is considered to be vorticity-free, or in other words, locally irrotational; i.e., 
\be \label{eq:aa14}
\tb{v}(x)=\frac{1}{m}\tilde{\na} \vartheta_{0}(x)\textcolor{black}{\equiv \frac{1}{m} \na \vartheta_{0}(x)},  
\ee
\textcolor{black}{see Eq.\eqref{eq:aa18} for the justification of $\tilde{\na} \to \na$ on $\vartheta_{0}(x)$ above}. Clearly, the classical mean-field phase $\vartheta_{0}(x)$ of the BEC state (see Eq.\eqref{eq:aa06}) now gets to act as the velocity potential\footnote{Though its is quite obvious from Eq.\eqref{eq:aa14}, we still write the velocity components as $ \tb{v}(x)=v_{r}(\tb{r})\hat{r}+v_{\theta}(\tb{r})\hat{\theta}+v_{\phi}(\tb{r})\hat{\phi}$ for sake of clarity, where, $v_{r}(\tb{r})=\frac{1}{m}\partial_{r} \vartheta_{0}(x)$, $v_{\theta}(\tb{r})=\frac{1}{m}\frac{\partial_{\theta} \vartheta_{0}(x)}{r}$ and $v_{\phi}(\tb{r})=\frac{1}{m}\frac{\partial_{\phi} \vartheta_{0}(x)}{r \sin\theta}$.} here in Eq.\eqref{eq:aa14}. 
\par 
This background velocity being irrotational plays a very crucial role in order to determine the metric (which eventually turns out to be \ti{Lorentzian} as seen by the phonons inside the fluid medium) of this particular $(3+1)$-d curved spacetime. In the present context, the fluid motion is assumed to be completely non-relativistic, i.e. $|\tb{v}|<<c$ where $c \approx 2.997 \times 10^{8} \hspace{0.05cm}$ ms$^{-1}$ is the speed of light in vacuum.
\par 
It is worth mentioning here that the local speed of sound\footnote{For reference, in the context of fluid dynamics, see eqn.(4.15) of \cite{Pitaevskii2003}.} inside the fluid medium is given by
\bea \label{eq:aa16}
c_{s}=\sqrt{n_{0}\mathsf{g}/m} \hspace{0.5cm},
\eea
\\
obviously, \hspace{0.2cm}$0 < c_{s} << c$ \hspace{0.2cm} in magnitude. \textcolor{black}{
If one starts by assuming the density $n_{0}$ to be position independent (which is pretty much the argument for considering a canonical acoustic black hole out of an incompressible and spherically symmetric fluid flow), due to the barotropic assumption, the pressure also becomes position independent. Thus for a barotropic fluid, $c_{s}$ becomes a position independent constant, refer to eqn.(25) of \cite{Barcelo2011}. In our analysis, the above Eq.\eqref{eq:aa16} effectively keeps $c_{s}$ as a constant all through. This is an approximation on our part in the present context. We make this approximation here because we don't need to consider the actual structure of the sonic horizon in that details. Obviously, when $c_{s}$ becomes position dependent, the sonic horizon would have a spread over space. But what we are basically concerned with here is analysing ``secondary'' waves (low frequency sonic modes) being generated outside the acoustic horizon by gaining energy from the ``primary'' waves (high frequency Hawking radiated sonic modes) due to quantum potential induced UV-IR coupling in a $(3+1)$-d curved spacetime. Hence, for that reason, we can safely regard $c_{s}$ to be a position independent constant and this approximation is legit.
}
\par
In terms of the velocity components, Eq.\eqref{eq:aa13} can now be rewritten as 
\begin{widetext}
\bea \nonumber
&\textcolor{black}{\tilde{\partial}_{t}} \hspace{0.05cm}\bigg[-\hat{\mathcal{A}} \hspace{0.1cm}
\Big(\textcolor{black}{\tilde{\partial}_{t}} \hspace{0.05cm}\{\vartheta_{2}(X)\vartheta_{1}(x)\} + \tb{v}(x) . \textcolor{black}{\tilde{\na}}\{\vartheta_{2}(X)\vartheta_{1}(x)\} \Big) \bigg]
\\ \label{eq:aa17} 
&+\textcolor{black}{\tilde{\na} }. \bigg(\bigg[-\hat{\mathcal{A}} \hspace{0.1cm}
\Big(\textcolor{black}{\tilde{\partial}_{t}} \hspace{0.05cm}\{\vartheta_{2}(X)\vartheta_{1}(x)\} + \tb{v}(x) . \textcolor{black}{\tilde{\na}}\{\vartheta_{2}(X)\vartheta_{1}(x)\} \Big) \bigg]\tb{v}(x) + \frac{n_{0}}{m}\textcolor{black}{\tilde{\na}} \{\vartheta_{2}(X)\vartheta_{1}(x) \}\bigg)=0, 
\eea
\end{widetext}
where, $\hat{\mathcal{A}}\equiv \mathsf{g}^{-1}\Big(1+\xi^{2} \tilde{\na}^{2}\Big)$, see Eq.\eqref{eq:aa12}.
\par
Let's note the following features of the above Eq.\eqref{eq:aa17}:
\begin{itemize}
\item
Essentially our choice of coordinate system is the Spherical polar coordinates throughout this paper.
\item
Each operation with $\hat{\mathcal{A}}$ naturally contains a Laplacian \textcolor{black}{which is considered to be at small scales ( i.e. $\tilde{\na}^{2}\rightarrow \na^{2} \equiv \partial^{\mathsf{j}}\partial_{\mathsf{j}}$ where $\mathsf{j}\equiv(r,\theta, \phi$) is the dummy index) }that comes with a pre-factor of $\xi^{2}$. But $\xi^{2}=\xi_{0}^{2}\hspace{0.05cm}\epsilon^{2}$, see Eq.\eqref{eq:aa10}. This fact is being stressed upon now, since we will be constructing a model here upto $\mathcal{O}(\epsilon^{2})$ accuracy. \textcolor{black}{Hence in the following prescription, we have to consider $\hat{\mathcal{A}}$ to be involving only the small scale derivatives and to keep it unperturbed because even the small scale Laplacian in $\hat{\mathcal{A}}$ already bears a $\epsilon^{2}$ pre-factor.}
\end{itemize}
\textcolor{black}{Now we explicitly mention the decomposition of the spacetime derivatives over independent \ti{multiple scales}} in order to separate out the dynamics upto $\mathcal{O}(\epsilon^{2})$. The multiple scale perturbation as considered here is defined by
\be \label{eq:aa18}
\textcolor{black}{\tilde{\partial}_{\mu} \rightarrow \partial_{\mu} + \epsilon \hspace{0.05cm} \partial_{\bar{\mu}} \hspace{0.2cm}},
\ee  
\textcolor{black}{where $\mu$ and $\bar{\mu}$ on r.h.s are merely the free indices subject to the restriction\footnote{It is quite redundant from Eq.\eqref{eq:aa18} that $\tilde{\partial}_{t} \to \partial_{t}+\epsilon \hspace{0.05cm}\partial_{\mathcal{T}}$, $\tilde{\partial}_{r} \to \partial_{r}+\epsilon \hspace{0.05cm}\partial_{R}$, $\tilde{\partial}_{\theta} \to \partial_{\theta}+\epsilon \hspace{0.05cm}\partial_{\Theta}$, $\tilde{\partial}_{\phi} \to \partial_{\phi}+\epsilon \hspace{0.05cm}\partial_{\Phi}$.} as mentioned previously under Eq.\eqref{eq:aa05}.}
\par
From now on, throughout the paper, 
\begin{itemize}
\item
by small-scale spacetime derivative, we would refer to this $\partial_{\mu} \equiv \frac{\partial}{\partial x^{\mu}}$ on r.h.s of Eq.\eqref{eq:aa18},
\item
and by large-scale spacetime derivative, we would refer to the above mentioned $\partial_{\bar{\mu}}\equiv \frac{\partial}{\partial X^{\bar{\mu}}}$.
\end{itemize}

\section{3. The Model: Analogue Gravity perspective}

In the following prescription, we'll be deriving our proposed model in detail where we show that the underlying non-relativistic BEC-system under consideration is very much capable of simulating the massive KG equation for some scalar field in a curved background. Since many elementary particles in nature do have non-zero mass, this happens to be a very significant and essential step, as already pointed out by Visser \ti{et al.} in 2005 \cite{Visser2005}, towards building some realistic analogue models rather than just providing some fictitious mathematical methodology.
\par 
In due course, we will deduce the explicit structure of the general acoustic metric in $(3+1)$-d and here we'll stick to the signature\footnote{In fact this convention of $(-,+,+,+)$ is clearly the reason that leads to generating a '$-$' sign in front of $\mathfrak{m}^{2}$ in Eq.\eqref{eq:aa21}.} of the metric tensor as $(-, +, +, +)$.

\subsection{3.1 \tu{Dynamics at different orders of $\epsilon$}}
Now Eq.\eqref{eq:aa18} is applied on Eq.\eqref{eq:aa17} and gives rise to a set of equations at different orders of $\epsilon$. The full model, considered till $\mathcal{O}(\epsilon^{2})$, is given by,
\bea \label{eq:aa19}
\mathcal{O}(1) &\Rightarrow & \hspace{0.1cm}\partial_{\mu}f^{\mu \nu} \partial_{\nu} \vartheta_{1}(x) =0, \\ \nonumber 
\\ \nonumber
\mathcal{O}(\epsilon) &\Rightarrow & \hspace{0.1cm}\bigg\lbrace \partial_{\mu} f_{1}^{\mu \bar{\mu}} \partial_{\bar{\mu}} + \partial_{\bar{\nu}} f_{1}^{\bar{\nu}\nu} \partial_{\nu} \bigg\rbrace \Big(\vartheta_{2}(X)\vartheta_{1}(x)\Big) =0, \\ \label{eq:aa20} \\ \nonumber 
\mbox{and,} \\ \label{eq:aa21}
 \mathcal{O}(\epsilon^{2}) &\Rightarrow & \hspace{0.1cm}\partial_{\bar{\mu}}f_{2}^{\bar{\mu} \bar{\nu}} \partial_{\bar{\nu}} \vartheta_{2}(X) - \mathfrak{m}^{2}\hspace{0.05cm} \vartheta_{2}(X) =0 ;
\eea
\\
where $[f^{\mu \nu}(x)]$, $[f_{1}^{\mu \bar{\mu}}(x, X)]$ or $[f_{1}^{\bar{\nu} \nu}(X, x)]$ and $[f_{2}^{\bar{\mu} \bar{\nu}}(X)]$ are all constructed as symmetric $4\times 4$ matrices explicitly written later. Here $\mathfrak{m}$ is the mass of the large length scale phonon modes, while it is strikingly found to be a finite function of the $\vartheta_{1}(x)$ field and thus Eq.\eqref{eq:aa21} is the \ti{massive} free KG equation for the field-amplitude $\vartheta_{2}(X)$. 
\par 
It is important to acknowledge the fact that, in the standard literature, one usually identifies this Eq.\eqref{eq:aa19} as the massless minimally coupled KG equation for a scalar field $\vartheta_{1}(x)$, see eqns.(248) and (254) of \cite{Barcelo2011}. But in our present framework, on top this usual massless picture at $\mathcal{O}(1)$, we come up with a massive KG equation in larger length scales at $\mathcal{O}(\epsilon^{2})$ subject to some constraint, given by Eq.\eqref{eq:aa20}, obtained at the intermediate $\mathcal{O}(\epsilon)$.
\par
In the beginning, the full expression of the mass term in Eq.\eqref{eq:aa21} contains a factor\footnote{This can be regarded as a coordinate-artifact due to the spherical polar coordinates. In cartesian coordinates, the formation of $\partial_{\bar{\mu}}f_{2}^{\bar{\mu} \bar{\nu}} \partial_{\bar{\nu}} \vartheta_{2}(X)$ while constructing Eq.\eqref{eq:aa21} would not have required a multiplication by any factor from left and hence, consequently, the large scale signature would have been completely absent in the expression of $\mathfrak{m}$. In our previous paper, see eqn.(23) of \cite{Sarkar2016}, we were able to construct a compact form to present the mass term because our formalism was in $(3+1)$-d Cartesian coordinates. } of $R^{2}\sin\Theta$. But through the process of scale reversion, as we are about to see in the next sub-section, this factor of $R^{2}\sin\Theta$ would obviously become $r^{2}\sin\theta$ giving rise to a re-scaled mass $\mathfrak{m}$ which gets inserted in the scale-reversed massive KG equation (i.e. Eq.\eqref{eq:aa25}) later. From now on, throughout the paper, whenever we speak about $\mathfrak{m}$, we would only refer to this re-scaled $\mathfrak{m}$ appearing in Eq.\eqref{eq:aa25}. The general expression of $\mathfrak{m}$, which was obtained using Mathematica 9.0 package, is extremely lengthy and hence not shown explicitly in this paper. However, after some physical approximations, a relatively tidier version is presented later by Eq.\eqref{eq:A1} in Appendix-I.
\par 
Later in Section 4.4, while deriving the full mass term, we'll talk about and clarify these steps one by one in detail. Now we are going to show the structures of the $f$-matrices to present our model. At $\mathcal{O}(1)$ in Eq.\eqref{eq:aa19}, the $[f^{\mu\nu}(x)]$ matrix is of the following form,
\bea \nonumber
 [f^{\mu\nu}] = \frac{r^{2}\sin\theta}{\mathsf{g}}\left( \begin{array}{cccc}
-1      &  -v_{r}                & -\frac{v_{\theta}}{r}           & -\frac{v_{\phi}}{r \sin\theta}                  
\\
\\
 -v_{r}    & \Big(c_{s}^{2}-v_{r}^{2}\Big)                 & -\frac{v_{r}v_{\theta}}{r}            & -\frac{v_{r}v_{\phi}}{r \sin\theta}                     
\\
\\
-\frac{v_{\theta}}{r}       & -\frac{v_{\theta}v_{r}}{r}                     & \frac{ \Big(c_{s}^{2}-v_{\theta}^{2}\Big)           }{r^{2}} & -\frac{v_{\theta}v_{\phi}}{r^{2}\sin\theta} 
\\ 
\\  
-\frac{v_{\phi}}{r \sin\theta}       & -\frac{v_{\phi}v_{r}}{r \sin\theta}                 & -\frac{v_{\phi}v_{\theta}}{r^{2} \sin\theta}            & \frac{\Big(c_{s}^{2}-v_{\phi}^{2}\Big)}{r^{2} \sin^{2}\theta}                 
\end{array} \right)_{.}\\ \label{eq:aa22}
\eea

In Eq.\eqref{eq:aa20}, $[f_{1}^{\mu \bar{\mu}}(x, X)]$ and $[f_{1}^{\bar{\nu} \nu}(X, x)]$ are the two matrices with their each and every corresponding entry being exactly the same, i.e.,
\\
\begin{footnotesize}
\begin{align} \nonumber
 & [f_{1}^{\mu \bar{\mu}}] \equiv  [f_{1}^{\bar{\nu} \nu}] \hspace{8cm} \\ \nonumber
 & =  \frac{R^{2} \sin\Theta \hspace{0.1cm} r^{2}\sin\theta}{\mathsf{g}} \left( \begin{array}{cccc}
-1      &  -v_{r}                & -\frac{v_{\theta}}{R}           & -\frac{v_{\phi}}{R \sin\Theta}                  
\\
\\
 -v_{r}    & \Big(c_{s}^{2}-v_{r}^{2}\Big)                 & -\frac{v_{r}v_{\theta}}{R}            & -\frac{v_{r}v_{\phi}}{R \sin\Theta}                     
\\
\\
-\frac{v_{\theta}}{r}       & -\frac{v_{\theta}v_{r}}{r}                     & \frac{\Big(c_{s}^{2}-v_{\theta}^{2}\Big)           }{R r}  & -\frac{v_{\theta}v_{\phi}}{R r \sin\Theta} 
\\ 
\\  
-\frac{v_{\phi}}{r \sin\theta}       & -\frac{v_{\phi}v_{r}}{r \sin\theta}                 & -\frac{v_{\phi}v_{\theta}}{R r \sin\theta}            & \frac{\Big(c_{s}^{2}-v_{\phi}^{2}\Big) }{R \sin\Theta \hspace{0.1cm} r \sin\theta}                
\end{array} \right)_{.}  \hspace{1cm}\\ \label{eq:aa23}
\end{align}
\end{footnotesize}

And finally the $[f_{2}^{\bar{\mu} \bar{\nu}}(X)]$ matrix, appearing in Eq.\eqref{eq:aa21}, is given by,
\begin{small}
\bea \nonumber
 [f_{2}^{\bar{\mu}\bar{\nu}}] = \frac{R^{2}\sin\Theta}{\mathsf{g}} \left( \begin{array}{cccc}
-1      &  -v_{r}                & -\frac{v_{\theta}}{R}           & -\frac{v_{\phi}}{R \sin\Theta}                  
\\
\\
 -v_{r}    & \Big(c_{s}^{2}-v_{r}^{2}\Big)                 & -\frac{v_{r}v_{\theta}}{R}            & -\frac{v_{r}v_{\phi}}{R \sin\Theta}                     
\\
\\
-\frac{v_{\theta}}{R}       & -\frac{v_{\theta}v_{r}}{R}                     & \frac{\Big(c_{s}^{2}-v_{\theta}^{2}\Big)}{R^{2}}            & -\frac{v_{\theta}v_{\phi}}{R^{2}\sin\Theta} 
\\ 
\\  
-\frac{v_{\phi}}{R \sin\Theta}       & -\frac{v_{\phi}v_{r}}{R \sin\Theta}                 & -\frac{v_{\phi}v_{\theta}}{R^{2} \sin\Theta}            & \frac{\Big(c_{s}^{2}-v_{\phi}^{2}\Big)}{R^{2} \sin^{2}\Theta}                 
\end{array} \right)_{.} \\ \label{eq:aa24}
\eea
\end{small}
Till this point, it pretty much sketches the basic introduction of our proposed model describing the dynamics of the phonon modes at different length scales and at different orders of the parameter $\epsilon$. Our main motive is to try investigating the massive KG equation found at $\mathcal{O}(\epsilon^{2})$ in detail through a simple mathematical framework.

\subsection{3.2 \tb{\tu{Scale reversion of the dynamics}}\hspace{0.2cm}($\partial_{\bar{\mu}} \to \frac{1}{\epsilon}\partial_{\mu}$)}

In our previous paper \cite{Sarkar2016}, we analyzed the situation in a flat background. But, for the large scale dynamics to see the curvature of spacetime, we must revert back systematically to the small length scales and this is exactly where our present analysis stands out to be very different from our previous analysis on flat spacetime.
\par 
On reversion back to the small scale dynamics from the large scales, each large scale space-time derivative does generate a factor of $1/\epsilon$ (i.e. $\partial_{\bar{\mu}} \rightarrow \frac{1}{\epsilon}\partial_{\mu}$) while undergoing the switching of scales. 

Till now, we are yet to talk about the constraint equation we found at $\mathcal{O}(\epsilon)$, see Eq.\eqref{eq:aa20}. First of all, we would start by identifying the construction of two matrices, viz. $[f_{1,\mathbb{I}}^{\mu \bar{\mu}}(x,X)]$ and $[f_{1,\mathbb{II}}^{\bar{\nu} \nu}](X,x)$, given by
\be \label{eq:aa73}
\Big(R^{2}\sin\Theta \times [f_{1,\mathbb{I}}^{\mu \bar{\mu}}]\Big) = [f_{1}^{\mu \bar{\mu}}] \equiv [f_{1}^{\bar{\nu} \nu}] = \Big(r^{2}\sin\theta \times [f_{1,\mathbb{II}}^{\bar{\nu} \nu}] \Big) \hspace{0.2cm},
\ee
with reference to Eq.\eqref{eq:aa23}. This readily gives the permit to rewrite Eq.\eqref{eq:aa20} in the following manner,
\begin{small}
\bea \nonumber
\bigg\lbrace \left(R^{2}\sin\Theta\right) \times \partial_{\mu}  f_{1,\mathbb{I}}^{\mu \bar{\mu}}\hspace{0.05cm}\partial_{\bar{\mu}} + \left(r^{2}\sin\theta \right) \times \partial_{\bar{\nu}}  f_{1,\mathbb{II}}^{\bar{\nu} \nu} \hspace{0.05cm}\partial_{\nu}\bigg\rbrace \Big(\vartheta_{2}(X)\vartheta_{1}(x)\Big) & \\ \nonumber
=0. &\\ \label{eq:aa74}
\eea
\end{small}
Through scale reversion, $R^{2}\sin\Theta$ written above naturally becomes $r^{2}\sin\theta \neq 0$,  $\partial_{\bar{\mu}} \rightarrow \frac{1}{\epsilon}\partial_{\mu}$ as just mentioned above and we find that both $[f_{1,\mathbb{I}}^{\mu \bar{\mu}}(x,X)]$ and $[f_{1,\mathbb{II}}^{\bar{\nu} \nu}(X,x)]$ strikingly take the form of $[f^{\mu \nu}(x)]$ as given by Eq.\eqref{eq:aa22}, the field-amplitude $\vartheta_{2}(X)$ gets scale transformed to give rise to a new complex scalar function, hence $\Big(\vartheta_{2}(X)\theta_{1}(x)\Big) \rightarrow \Xi(x)$. Thus Eq.\eqref{eq:aa74} is rewritten in the scale-reversed form as,
\be \label{eq:aa75}
 \partial_{\mu} f^{\mu \nu} \partial_{\nu} \hspace{0.05cm} \hspace{0.1cm} \Xi(x)=0.
\ee
Hence we come up with exactly the same dynamics for $\Xi(x)$ at $\mathcal{O}(\epsilon)$ as we had obtained for only $\vartheta_{1}(x)$ at $\mathcal{O}(1)$, see Eq.\eqref{eq:aa19}. This clearly indicates that we are not going to get anything new at this stage because $\mathcal{O}(\epsilon)$-dynamics practically captures the same field over small length scales. \textcolor{black}{This happens particularly because there is no source term at $\mathcal{O}(\epsilon)$.}
\par 
So we have to move on to analyzing the next order, i.e. $\mathcal{O}(\epsilon^{2})$, dynamics to see the new structure in the field after having it reverted back to the small scales. By inspection, its quite evident that $[f_{2}^{\bar{\mu} \bar{\nu}}(X)]$ in Eq.\eqref{eq:aa21} would again take the form exactly as $[f^{\mu \nu}(x)]$ on switching back to the small scales and $\vartheta_{2}(X) \rightarrow \varphi(x)$ as mentioned already. Therefore Eq.\eqref{eq:aa21} gets scale transformed as the following,
\be \label{eq:aa25}
\partial_{\mu}f^{\mu \nu} \partial_{\nu} \varphi(x) - \epsilon^{2}\mathfrak{m}^{2}\hspace{0.05cm} \varphi(x) =0 ,
\ee
where $[f^{\mu \nu}]$ is already given by Eq.\eqref{eq:aa22}. 
\par
Eq.\eqref{eq:aa25} is purposely multiplied by a real scalar constant $\mathsf{g}/c_{s}$ and gives rise to,
\be \label{eq:aa26}
\partial_{\mu}f_{\mbox{\begin{tiny}New\end{tiny}}}^{\mu \nu} \partial_{\nu} \varphi(x) - \frac{\mathsf{g}}{c_{s}}\epsilon^{2}\mathfrak{m}^{2}\hspace{0.05cm} \varphi(x) =0 , 
\ee
\hspace{0.5cm}where the contravariant $f$-matrix now being scaled up as
\bea \nonumber
& f_{\mbox{\begin{tiny}New\end{tiny}}}^{\mu \nu}= \frac{\mathsf{g}}{c_{s}}[f^{\mu\nu}] \hspace{6cm}\\ \nonumber \\ \nonumber 
&\hspace{0.5cm}=  \frac{r^{2}\sin\theta}{c_{s}}\left( \begin{array}{cccc}
-1      &  -v_{r}                & -\frac{v_{\theta}}{r}           & -\frac{v_{\phi}}{r \sin\theta}                  
\\
\\
 -v_{r}    & \Big(c_{s}^{2}-v_{r}^{2}\Big)                 & -\frac{v_{r}v_{\theta}}{r}            & -\frac{v_{r}v_{\phi}}{r \sin\theta}                     
\\
\\
-\frac{v_{\theta}}{r}       & -\frac{v_{\theta}v_{r}}{r}                     & \frac{ \Big(c_{s}^{2}-v_{\theta}^{2}\Big)           }{r^{2}} & -\frac{v_{\theta}v_{\phi}}{r^{2}\sin\theta} 
\\ 
\\  
-\frac{v_{\phi}}{r \sin\theta}       & -\frac{v_{\phi}v_{r}}{r \sin\theta}                 & -\frac{v_{\phi}v_{\theta}}{r^{2} \sin\theta}            & \frac{\Big(c_{s}^{2}-v_{\phi}^{2}\Big)}{r^{2} \sin^{2}\theta}                 
\end{array} \right)_{.} \\ \label{eq:aa27}
\eea
This process of scale reversion is actually a very important step here. Had we not switched back to the small scales, the spacetime metric arising out of $[f_{2}^{\bar{\mu}\bar{\nu}}(X)]$ corresponding to the large scale dynamics would have defined the background geometry to be effectively flat because its entries are basically the velocity field components all of which were retained at small scales (see Eq.\eqref{eq:aa24}) and thus act as just constants w.r.t the large scale spacetime derivatives. 

\subsection{3.3 \tu{The covariant massive KG equation}}

In order to cast the above Eq.\eqref{eq:aa26} into a Lorentz invariant form, it is required to identify the corresponding covariant structure and hence the introduction of the effective metric (or in other words, acoustic metric) in place of the respective $f$-matrix is essential. From this point onward, we talk about the \ti{covariant} massive minimally coupled free KG equation.
\par
Say, $[g_{\mu \nu}(x)]$ be the general acoustic metric that actually defines the $(3+1)$-d curved spacetime under consideration with its determinant, given by, $g=\text{det}[g_{\mu \nu}(x)]$. Considering Eq.\eqref{eq:aa26}, one identifies 
\be \label{eq:aa76}
f_{\mbox{\begin{tiny}New\end{tiny}}}^{\mu \nu}=\sqrt{|g|}g^{\mu \nu},
\ee
\bea \nonumber
\Rightarrow \hspace{0.5cm}\text{det}[f_{\mbox{\begin{tiny}New\end{tiny}}}^{\mu \nu}] = \text{det}[\sqrt{|g|}g^{\mu \nu}] = (\sqrt{|g|})^{4} \hspace{0.05cm} \text{det}[g^{\mu \nu}] \\ \nonumber
= (\sqrt{|g|})^{4} \hspace{0.05cm} g^{-1} = g. \\ \nonumber
\because \text{det}[f_{\mbox{\begin{tiny}New\end{tiny}}}^{\mu \nu}] = -c_{s}^{2}r^{4}\sin^{2}\theta \hspace{0.1cm},\hspace{0.5cm} \therefore \hspace{0.1cm}g= -c_{s}^{2}r^{4}\sin^{2}\theta, \\ \nonumber
\mbox{and obviously}, \hspace{0.5cm} g^{\mu \nu}=\frac{1}{\sqrt{|g|}}f_{\mbox{\begin{tiny}New\end{tiny}}}^{\mu \nu} = \frac{1}{c_{s}r^{2}\sin\theta}f_{\mbox{\begin{tiny}New\end{tiny}}}^{\mu \nu}. \\ \label{eq:aa28}
\eea 
Now, from Eq.\eqref{eq:aa28}, it is just trivial to find the acoustic metric which is of the following form,
\bea \nonumber
[g_{\mu\nu}] = \left( \begin{array}{cccc}
-\Big(c_{s}^{2}-\tb{v}^2 \Big)      &  -v_{r}                & -v_{\theta}r          & -v_{\phi}r \sin\theta                
\\
\\
 -v_{r}    & 1              & 0            & 0                     
\\
\\
-r v_{\theta}       & 0                     & r^{2}            & 0 
\\ 
\\  
-r \sin\theta \hspace{0.02cm}v_{\phi}       & 0               & 0            & r^{2}\sin^{2}\theta                 
\end{array} \right)_{.}\\ \label{eq:aa29}
\eea
It should be noted that, in general relativity, the spacetime metric (which does bear the feature of the background geometry) is related to the distribution of matter (i.e. the stress-energy tensor) through the Einstein's-Field-Equations; whereas, the acoustic metric $[g_{\mu\nu}(x)]$ here happens to be related to the background velocity field \textcolor{black}{($\tb{v(\tb{r})}$)} \textcolor{black}{as well as the local speed of sound ($c_{s}$)} in a way more simpler algebraic fashion. Some striking features of this $[g_{\mu\nu}(x)]$ from topological aspect and regarding 'stable causality' have been discussed by Visser in \cite{Visser1998}-pp.1773-1774.
\par 
Finally, Eq.\eqref{eq:aa26} is re-written in the standard covariant form, given by,
\bea \nonumber
&\frac{1}{\sqrt{|g|}} \partial_{\mu}\left(\sqrt{|g|}g^{\mu \nu} \partial_{\nu}\right) \varphi(x) - \frac{1}{\sqrt{|g|}}\frac{\mathsf{g}}{c_{s}}\epsilon^{2}\mathfrak{m}^{2}\hspace{0.05cm} \varphi(x) =0, \\ \label{eq:aa30}
& \mbox{i.e.} \hspace{0.5cm} \Big(\nabla_{\mu}\nabla^{\mu} - \mathcal{M}^{2}\Big)\hspace{0.05cm} \varphi(x) =0,
\eea
where $\nabla_{\mu}$ is obviously the covariant derivative and the final mass-term $\mathcal{M}$ is determined through 
\be \label{eq:aa31}
\mathcal{M}^{2}=\frac{1}{\sqrt{|g|}}\frac{\mathsf{g}}{c_{s}}\epsilon^{2}\mathfrak{m}^{2},
\ee
where $g=-c_{s}^{2}r^{4}\sin^{2}\theta$ \hspace{0.2cm}in our present model (refer to Eq.\eqref{eq:aa28}). 
\par
One thing is to be understood that if the mass term is dropped (i.e. $\mathcal{M}=0$) off from the solution of the scalar function $\varphi(x)$ as determined through Eq.\eqref{eq:aa30} (or in other words, Eq.\eqref{eq:aa25}) - then what we come up with as a solution is nothing but $\vartheta_{1}(x)$ via Eq.\eqref{eq:aa19}.

\section{4. The \ti{canonical} acoustic black hole}

With a strong motive to examine how closely the acoustic metric can get to mimic the standard Schwarzschild geometry in gravity, one usually considers some specific symmetry in the analogue spacetime to move ahead. If one starts by considering an analogue gravity scenario in a spherically symmetric flow (we'll consider the flow to be non-relativistic here) of a barotropic incompressible inviscid fluid, one comes up with a solution called \ti{canonical acoustic black hole} found by Visser \cite{Visser1998} in 1998.
\par 
In principle, we would restrict ourselves only to the stationary\footnote{\textcolor{black}{Stationary solutions are of special interest and significance because they are regarded as the ``end states'' of a gravitational collapse.}}, non-rotating, \textcolor{black}{asymptotically flat} canonical acoustic black holes. Thus, in our following prescription, \textcolor{black}{the notions of \ti{apparent} and \ti{event horizons (acoustic)}}\footnote{\textcolor{black}{The event horizon is a global feature, it could be difficult to actually locate such a boundary while being handed with a metric in an arbitrary set of coordinates. Usually it is defined to be the boundary of the region from which even the null geodesics can not escape - strictly speaking, this is \ti{future} event horizon.}} coincide and the distinction becomes immaterial. \textcolor{black}{In the language of the standard general relativity, an event horizon is a null hypersurface that separates those spacetime points connected to infinity through timelike path from those that are not \cite{Carroll2004}. }
\par
\textcolor{black}{Since a canonical acoustic BH, \textcolor{black}{as considered here}, is indeed stationary and 
asymptotically flat, every event horizon is a Killing horizon for some Killing vector field, say $\sigma^{\mu}$. Due to the time-translational and axial symmetry of the metric (refer to Eq.\eqref{eq:aa36} later), obviously there are two Killing vector fields, viz. $\sigma_{(t)}\equiv \partial_{t}$ and $\sigma_{(\phi)}\equiv \partial_{\phi}$ which go from timelike to spacelike \textcolor{black}{and vice-versa} at the event horizon. \ti{Killing horizon} is formally defined to be the null hypersurface on which the Killing vector field becomes null.}
\par
\textcolor{black}{The acoustic (Killing) horizon is formed once the radial component of the background fluid-velocity ($v_{r}$) exceeds the local speed of sound ($c_{s}$), refer to eqn.(45) of \cite{Visser1998}. }
\subsection{4.1 \tu{Massive KG equation in the canonical spacetime}}
Since, the classical mean-field $n_{0}$ and $\vartheta_{0}(x)$ have got to satisfy the continuity equation (Eq.\eqref{eq:aa03}), so clearly,
\bea \nonumber
0=\partial_{t}\hspace{0.05cm}n_{0}+\frac{1}{m}\na.\Big(n_{0}\na \vartheta_{0}(x)\Big)=\textcolor{black}{n_{0}}\na.\tb{v}(x),& \\ \label{eq:aa32}
\Rightarrow |\tb{v}(x)| \propto \frac{1}{r^{2}}.&
\eea
 And thus, through a normalization constant finite $r_{\text{\tiny 0}}> 0$, the background velocity-field
\footnote{Refer to eqn.(54) of \cite{Visser1998}.} is set to be 
\be \label{eq:aa33}
|\tb{v}(\tb{r})|=v_{r}=c_{s}\frac{r_{\text{\tiny 0}}^{2}}{r^{2}}, \hspace{0.5cm}(\forall \hspace{0.1cm} 0< r <\infty).
\ee
\par
Considering $v_{\theta}=0=v_{\phi}$, Eq.\eqref{eq:aa29} gives rise to the exact acoustic metric that describes the present scenario. The line element is given by, 
\be \label{eq:aa34}
ds^{2}=-c_{s}^{2}dt^{2}+\Big( dr \pm v_{r} \hspace{0.05cm}dt \Big)^{2}+r^{2}\Big(d\theta^{2}+\sin^{2}\theta \hspace{0.05cm}d\phi^{2}\Big).
\ee
It is to be noted that when $v_{r} > 0$, which is also the usual convention, then it would be a `$-$' sign in front of $v_{r}dt$ in the second term on r.h.s of the above Eq.\eqref{eq:aa34}. Otherwise, it would be a `$+$' sign over there when $v_{r} < 0$, i.e. when the fluid flow is considered to be in the opposite direction.
\par
Instead of the laboratory time $t$, one can now introduce the analogue-Schwarzschild-time coordinate $\tau$ via the simple coordinate transformation as
\be \label{eq:aa35}
t \longrightarrow \tau = t \mp \bigg(\frac{r_{\text{\tiny 0}}}{2c_{s}}\tan^{-1}(r/r_{\text{\tiny 0}})+\frac{r_{\text{\tiny 0}}}{4c_{s}}\hspace{0.05cm}\ln\left|\frac{1-\frac{r}{r_{\text{\tiny 0}}}}{1+\frac{r}{r_{\text{\tiny 0}}}}\right| \bigg),
\ee
and, using Eq.\eqref{eq:aa33} and Eq.\eqref{eq:aa34}, it readily gives rise to a somewhat "Schwarzschild-like" line element\footnote{Refer to eqn.(56) of \cite{Vieira2016}} describing a canonical acoustic black hole, given by, 
\begin{align} \nonumber
 & ds^{2}=-\frac{c_{s}^{2}}{r^{4}}\Delta(r)\hspace{0.05cm} d\tau^{2}\hspace{0.1cm}+\hspace{0.1cm}\frac{r^{4}}{\Delta(r)}dr^{2}\hspace{0.1cm}+\hspace{0.1cm}r^{2}\Big(d\theta^{2}\hspace{0.1cm}+\hspace{0.1cm}\sin^{2}\theta \hspace{0.05cm}d\phi^{2}\Big), \\ \label{eq:aa36}
 &\hspace{4cm}\text{where}, \hspace{0.5cm} \Delta(r)=r^{4}-r_{\text{\tiny 0}}^{4} \hspace{0.1cm}. 
\end{align}
It is redundant to read off the acoustic metric from the above line element (Eq.\eqref{eq:aa36}) as the following,
\be \label{eq:aa37}
[g_{\mu\nu}]_{\text{Canonical BH}} \equiv \left( -\frac{c_{s}^{2}}{r^{4}}\Delta(r), \hspace{0.05cm}\frac{r^{4}}{\Delta(r)}, \hspace{0.05cm}r^{2}, \hspace{0.05cm} r^{2}\sin^{2}\theta \right),
\ee
and we see that the spacetime of a canonical acoustic BH is asymptotically flat \textcolor{black}{and naturally has a ``physical singularity'' at $r=0$, which is again quite obvious from Eq.\eqref{eq:aa33} - the background fluid velocity diverges at the center of the canonical acoustic BH}. Evidently, $r_{\text{\tiny 0}}$ is the \textcolor{black}{Killing} horizon (or, \ti{sonic horizon} to be more precise) of the canonical acoustic black hole. As far as the physical picture is concerned, beyond this point $r=r_{\text{\tiny 0}}$\hspace{0.05cm}, the fluid essentially becomes supersonic w.r.t an observer sitting at some large $r\rightarrow \infty$; i.e. $v\geqslant c_{s}$ holds true $\forall \hspace{0.05cm}r\leqslant r_{\text{\tiny 0}}$ which is again quite obvious from Eq.\eqref{eq:aa33} as well.
\par
Through the acoustic metric $[g_{\mu\nu}(x)]_{\text{Canonical BH}}$ \hspace{0.05cm}in Eq.\eqref{eq:aa37}, the covariant massive KG equation (Eq.\eqref{eq:aa30}), in the spacetime of a canonical acoustic black hole, boils down to the following form 
\bea \nonumber 
&-&\frac{r^{4}}{c_{s}^{2} \Delta(r)}\partial_{\begin{tiny}\tau\tau\end{tiny}}\hspace{0.08cm}\varphi(\tau, \tb{r}) \hspace{0.1cm}+\hspace{0.1cm}\frac{1}{r^{2}}\partial_{r}\left( \frac{\Delta(r)}{r^{2}}\partial_{r}\hspace{0.08cm}\varphi(\tau, \tb{r}) \right) \\ \nonumber 
&+& \frac{1}{r^{2}\sin\theta}\partial_{\theta}\Big(\sin\theta \hspace{0.05cm}\partial_{\theta}\hspace{0.08cm}\varphi(\tau, \tb{r}) \Big)+\frac{1}{r^{2} \sin^{2}\theta}\partial_{\phi \phi}\hspace{0.08cm}\varphi(\tau, \tb{r}) \\ \nonumber 
&&\hspace{5cm}- \mathcal{M}^{2}\hspace{0.08cm}\varphi(\tau, \tb{r}) =0. \\ \label{eq:aa38} 
\eea
The above Eq.\eqref{eq:aa38} has some important features:
\begin{itemize}
\item
The spacetime given by Eq.\eqref{eq:aa36} is clearly \ti{static} and has a time-translational symmetry. Thus the temporal part of $\varphi(\tau, \tb{r})$ that solves the above differential equation (Eq.\eqref{eq:aa38}) can easily be separated out as $e^{-i\omega \tau}$ \hspace{0.1cm} ($\forall \hspace{0.1cm} 0<\omega<\infty$), where $\omega$ is the frequency (or, equivalently \ti{energy} in $\hbar=1$ unit) of the particles associated to the $\varphi(\tau, \tb{r})$ field.
\item
$[g_{\mu\nu}(x)]_{\text{Canonical BH}}$ \hspace{0.05cm} describes a spacetime that also has a rotational invariance with respect to $\phi$ and similarly the azimuthal part of the solution to Eq.\eqref{eq:aa38} is obviously $e^{i \mathsf{m}\phi}$, where $\mathsf{m}=\pm 1,\pm 2,\pm 3,...$ is the azimuthal quantum number.
\item
From Eq.\eqref{eq:aa38}, its evident that the general angular solution can be given in terms of the standard spherical harmonics,
\be \label{eq:aa39}
\mathcal{Y}^{l}_{\mathsf{m}}(\theta, \phi)=\mathcal{P}^{l}_{\mathsf{m}}(\cos\theta)e^{i\mathsf{m}\phi},
\ee
where $\mathcal{P}^{l}_{\mathsf{m}}$-s are obviously the Legendre polynomials with $l$ being an integer such that $|\mathsf{m}|\leqslant l$.
\end{itemize}

\subsection{4.2 \tu{The Radial Solution}}
Therefore, to solve Eq.\eqref{eq:aa38}, we can consider the following ansatz as
\be \label{eq:aa40}
\varphi(\tau, \tb{r})=\frac{1}{r}\mathcal{R}(r)\mathcal{Y}^{l}_{\mathsf{m}}(\theta, \phi)e^{-i\omega \tau},
\ee
where $\mathcal{R}(r)$ is just the radial function to be determined.
\par
Substituting Eq.\eqref{eq:aa40} back into Eq.\eqref{eq:aa38}, we find that
\bea \nonumber
&& \frac{1}{r^{2}}\frac{d}{dr}\left(\frac{\Delta(r)}{r^{2}} \frac{d}{dr}\left(\frac{\mathcal{R}(r)}{r} \right) \right) \\ \label{eq:aa41}
&& \hspace{0.5cm}+\left[\frac{\omega^{2}r^{4}}{c_{s}^{2}\Delta(r)} - \left(\mathcal{M}^{2}+\frac{l(l+1)}{r^{2}} \right)\right] \frac{\mathcal{R}(r)}{r} =0.
\eea
Now this has become a linear second-order ordinary differential equation in $r$ for an undetermined function $\mathcal{R}(r)$. We go on reducing Eq.\eqref{eq:aa41} further to check the singularity (if any) at various points, because in order to solve the radial differential equation, we are about to pick the Frobenius ansatz for $\mathcal{R}(r)$ and adopt the method of series solution.
\par 
By inspection, we find the nature of singularities (for detailed explanations, see Appendix-II) of the above differential equation (Eq.\eqref{eq:aa41}) as the following
\begin{itemize}
\item 
$r=0$ is a \ti{regular} singular point (\textcolor{black}{It is to be understood clearly that $r=0$ is indeed a point of physical singularity for the metric (Eq.\eqref{eq:aa37}) itself, but not for the above ordinary differential equation (Eq.\eqref{eq:aa41}).  For Eq.\eqref{eq:aa41}, $r=0$ is a regular or removable singular point. Refer to \ti{Appendix - II} for a detailed argument}.) .
\item
$r=r_{\text{\tiny 0}}$ is also a \ti{regular} singular point.
\item 
And the point $r\to \infty$ is an \ti{irregular} singular point.
\end{itemize}
 
We introduce a new coordinate \hspace{0.02cm}$\chi$ \hspace{0.02cm} in order to simplify the structure of the above Eq.\eqref{eq:aa41}. The coordinate transformation, actually known as the \ti{Eddington-Finkelstein} tortoise coordinates (also known as \ti{Regge-Wheeler} coordinates), basically allows one to use the new coordinate \hspace{0.02cm}$\chi$ \hspace{0.02cm} even in the interior region of the acoustic black hole (i.e., when $r < r_{\text{\tiny 0}}$). 
\par
In the present context\footnote{In case of the usual Schwarzschild metric in gravity, this tortoise coordinate becomes $\chi=r+r_{s}\hspace{0.01cm}\ln\left|\frac{r}{r_{s}}-1 \right|$ where $r_{s}$ is the Schwarzschild radius. See eqn.(5.108) of \cite{Carroll2004}.}, the coordinate transformation is given by, 
\begin{small}
\be \label{eq:aa43}
r \longrightarrow \chi\equiv \chi(r)=\pm \frac{r}{c_{s}} \mp \hspace{0.05cm}\frac{r_{\text{\tiny 0}}}{2c_{s}}\tan^{-1}\left(\frac{r}{r_{\text{\tiny 0}}}\right) \hspace{0.1cm}\pm \frac{r_{\text{\tiny 0}}}{4c_{s}}\hspace{0.05cm}\ln\left|\frac{1-\frac{r}{r_{\text{\tiny 0}}}}{1+\frac{r}{r_{\text{\tiny 0}}}} \right| .
\ee
\end{small}
Conventionally, \\
$\chi=+ \frac{r}{c_{s}} - \hspace{0.05cm}\frac{r_{\text{\tiny 0}}}{2c_{s}}\tan^{-1}\left(\frac{r}{r_{\text{\tiny 0}}}\right) \hspace{0.1cm} + \frac{r_{\text{\tiny 0}}}{4c_{s}}\hspace{0.05cm}\ln\left|\frac{1-\frac{r}{r_{\text{\tiny 0}}}}{1+\frac{r}{r_{\text{\tiny 0}}}} \right| $ and its evident that 
\begin{itemize}
\item 
\hspace{0.02cm}$\chi$ \hspace{0.02cm} approaches $0$ as $r\to 0$,
\item
\hspace{0.02cm}$\chi$ \hspace{0.02cm} approaches $-\infty$ as $r\to r_{\text{\tiny 0}}$ from the either sides of the \textcolor{black}{acoustic Killing horizon},
\item 
and, as $r\to +\infty$, \hspace{0.02cm}$\chi$ \hspace{0.02cm} approaches $+\infty$ . 
\end{itemize}
Hence in the exterior region of the acoustic black hole, i.e. $\forall \hspace{0.1cm}r_{\text{\tiny 0}} < r < +\infty$, \hspace{0.02cm}$\chi$ \hspace{0.02cm} is found to be continuous: $-\infty < \chi < +\infty$.
\par
The tortoise coordinate is intended to grow infinite at the appropriate rate such as to cancel out the singular behaviour of the spacetime at $r=r_{\text{\tiny 0}}$ (the coordinate-singularity is quite vivid from the Eq.\eqref{eq:aa37}) which is essentially nothing but the artifact of the choice of coordinates.
\\
\par
Via the above transformation described in Eq.\eqref{eq:aa43}, one can easily reduce Eq.\eqref{eq:aa41} to the following form,
\begin{small}
\bea \nonumber
&& \frac{d^{2} \mathcal{R}(r)}{d\chi^{2}}+\left[\omega^{2} - \left(\mathcal{M}^{2}+\frac{\bar{l}}{r^{2}} +\frac{4r_{\text{\tiny 0}}^{4}}{r^{6}}\right)c_{s}^{2}\left(1-\frac{r_{\text{\tiny 0}}^{4}}{r^{4}}\right)\right] \mathcal{R}(r) =0, \\ \nonumber \\ \label{eq:aa44}
&& \text{where,} \hspace{0.5cm}\bar{l}\equiv l(l+1). 
\eea
\end{small}
\par
Our aim is to find a series solution of the above Eq.\eqref{eq:aa44} valid in the exterior region of the space-time; i.e. $\forall \hspace{0.04cm}r > r_{\text{\tiny 0}}$. We pick an ansatz, as chosen by Elizalde \cite{Elizalde1988}, of the following form,
\begin{align} \label{eq:aa45}
& \mathcal{R}(r)=\alpha \hspace{0.05cm} e^{\pm i\hspace{0.05cm}\left[\Bbbk \hspace{0.05cm} \chi \hspace{0.1cm} +\hspace{0.1cm}  h(\rho) \right]}\hspace{0.05cm};& \\ \nonumber \mbox{where}\\ \nonumber
& \rho=1-\frac{r_{\text{\tiny 0}}}{r}, \hspace{0.3cm} h(\rho)=\beta \hspace{0.05cm}\ln(1-\rho)+\sum_{\mathclap{n=1}}^{\infty}\left(a_{n}\hspace{0.05cm}\rho^{n}\right),& \\ \label{eq:aa46}
& \hspace{4cm}\text{necessarily, $a_{1}\neq 0$}.&
\end{align}

Here $\alpha$ is any arbitrary constant while $\Bbbk$ in Eq.\eqref{eq:aa45} and $\beta$ in Eq.\eqref{eq:aa46} are the constants to be determined. One should note that all the $a_{n}$\hspace{0.05cm}-s above are nothing but the Frobenius-coefficients.
\par
In order to avoid any conflict of notations, we would like to clearly mention here that this $n$ in Eq.\eqref{eq:aa46} is simply the dummy index of the infinite sum and has nothing to do with the $n$ which was introduced previously as density in Eq.\eqref{eq:aa02}. From now on, we would only consider the $+$ sign in front of $i$ on r.h.s of Eq.\eqref{eq:aa45}; but one should notice that an ansatz with just $-i$ over there would also do equally.
\par
Clearly in the exterior region, i.e. $\forall \hspace{0.08cm}r\geqslant r_{\text{\tiny 0}}$, we always have $0\leqslant \rho \leqslant 1$ (see Eq.\eqref{eq:aa46}) because 
\begin{align} \label{eq:aa47}
\lim_{r\to r_{\text{\tiny 0}}} \rho = 0 \hspace{0.3cm}; \hspace{0.3cm}\lim_{r\to \infty} \rho = 1. 
\end{align}
This is exactly what justifies the form of $h(\rho)$ as considered in Eq.\eqref{eq:aa46} to be legit in the exterior region.
\par
Most of the tedious algebraic expressions are explicitly shown in the Appendix, and we will be sketching only the important steps here. Inserting Eqs.\eqref{eq:aa45} and \eqref{eq:aa46} back into Eq.\eqref{eq:aa44}, followed by further simplifications (see Appendix-III, Eqs.\eqref{eq:A7}, \eqref{eq:A8} for details), one can rewrite Eq.\eqref{eq:aa44} in terms of the variable $\rho$ as the following,
\begin{widetext}
\begin{align} \nonumber
&\left(-\frac{1}{r_{\text{\tiny 0}}^{2}}\right)(1-\rho )^{10} \times \hspace{0.05cm} \Bigg[\frac{\Bbbk^2
   r_{\text{\tiny 0}}^2}{(1-\rho)^{10}}+\frac{2 \beta  \Bbbk \left(\rho ^3-4 \rho ^2+6 \rho -4\right) \rho  r_{\text{\tiny 0}} c_s}{(1-\rho)^9}  +\beta c_s^2 \left\lbrace\frac{1}{(1-\rho)^4}-1\right\rbrace \left\lbrace\frac{\beta -i}{(1-\rho )^4}-\beta +5 i\right\rbrace \\ \nonumber
&+\sum_{\mathclap{n=1}}^{\infty}\bigg\lbrace \frac{n^2 \left(\rho ^3-4 \rho ^2+6 \rho -4\right)^2 a_n^2 c_s^2 \boxed{\rho ^{2 n}}}{(1-\rho)^6} -\frac{1}{(1-\rho)^8}\bigg( i c_s n a_n  \boxed{\rho^{n}}  \left(\rho ^3-4 \rho ^2 +6 \rho -4\right) \Big( -c_s (1-\rho) \Big[ n \big(\rho ^4-5 \rho ^3+10 \rho ^2 \\ \nonumber 
& \hspace{16.5cm} -10 \rho +4 \big) \\ \nonumber
&  +\rho  \Big(2 i \beta \left(\rho ^3-4 \rho ^2+6 \rho -4\right) +5 \rho ^3-19 \rho ^2+26 \rho -14\Big)\Big]  -2 i \Bbbk r_0\Big)\bigg) \bigg\rbrace \Bigg] \\ \label{eq:aa51}
& + \Bigg[\omega ^2 + \frac{1}{r_{\text{\tiny 0}}^2}\left \lbrace c_s^2 \rho  (\rho -1)^2 \left(\rho ^3-4 \rho ^2+6 \rho -4\right) \left(\bar{l}+4(\rho -1)^4\right)+\mathcal{M}^2 \rho  \left(\rho ^3-4 \rho ^2+6 \rho -4\right)
   r_{\text{\tiny 0}}^2 c_s^2 \right\rbrace \Bigg] = 0.
\end{align}
\end{widetext}
By clubbing the corresponding coefficients of the various powers of $\rho$ from the first square-bracket of the above equation, one keeps all of them on the left-hand-side; while the second square-bracket of the above equation is giving rise to the $n$-independent terms, all of which are moved to the right-hand-side. Thus the above Eq.\eqref{eq:aa51} can be neatly re-written as Eq.\eqref{eq:A9}, see Appendix-III. 
\par 
In order to find the recursion relation, this Eq.\eqref{eq:A9} can now be compacted as, 

\bea \nonumber
&\text{l.h.s of Eq.\eqref{eq:A9}}= \underbrace{\sum_{\mathclap{n=1}}^{\infty}\sum_{\mathclap{k=0}}^{10}\rho^{n+k} \hspace{0.1cm}\mathfrak{f}_{k}^{\mathbb{I}}(n)}_{\mathcal{S}_{1} \hspace{0.2cm}(\text{say})}+\underbrace{\sum_{\mathclap{n=1}}^{\infty}\sum_{\mathclap{p=0}}^{10}\rho^{2n+p} \hspace{0.1cm}\mathfrak{f}_{p}^{\mathbb{II}}(n)}_{\mathcal{S}_{2} \hspace{0.2cm}(\text{say})} \\ \nonumber &\Updownarrow \hspace{7cm} \\ \nonumber
&\text{r.h.s of Eq.\eqref{eq:A9}}= \mathcal{F}(\rho)\equiv F_{0}\rho^{0}+F_{1}\rho^{1}+...+F_{11}\rho^{11} \hspace{0.2cm},\\ \label{eq:aa53}
\eea
having identified the respective coefficient(s) of each power of $\rho$ on both sides by some specified functions, given by, $\mathfrak{f}_{k}^{\mathbb{I}}(n)$ ($\forall \hspace{0.05cm}k$) and $\mathfrak{f}_{p}^{\mathbb{II}}(n)$ ($\forall \hspace{0.05cm}p$) are obviously defined in consistence with their corresponding explicit forms written in the l.h.s of Eq.\eqref{eq:A9}. On the other hand, the $F_{0}$, $F_{1}$,..., $F_{11}$, as described above in Eq.\eqref{eq:aa53}, are all independent of $n$. These are the respective coefficients of $\rho^{0}$, $\rho^{1}$,..., $\rho^{11}$ in the full source term $ \mathcal{F}(\rho)$.
\par
Our motive is to exhaust each and every term of $\mathcal{F}(\rho)$ by the corresponding term(s) picked from $\mathcal{S}_{1}$ and $\mathcal{S}_{2}$ via the power-matching of $\rho$ on both sides of Eq.\eqref{eq:A9} and then try investigating the \ti{recursion relation}. For the sake of clarity and lucidness, a vast part of the calculation\footnote{The calculation being extremely tedious, a part of it has been worked out via Mathematica-9.0 package, however the key steps are mentioned systematically.} is shown in details and step by step in the Appendix section, see Appendix-III, towards the end of the paper.
\par
At this point, we need to refer to the Eqs.\eqref{eq:A12}, \eqref{eq:A15}, \eqref{eq:A17} and \eqref{eq:A19} from Appendix-III and having these equations clubbed together, one can rewrite Eq.\eqref{eq:aa53} in the following manner,
\begin{widetext}
\begin{footnotesize}
\begin{align} \nonumber
 \because \hspace{0.3cm} \left(\mathcal{S}_{1}+\mathcal{S}_{2} \right) &=\mathcal{F}(\rho)  \\ \nonumber
  \Rightarrow \hspace{0.3cm}\sum_{\mathclap{n=1}}^{12-k-1}\sum_{\mathclap{k=0}}^{10}\rho^{n+k} \hspace{0.1cm}\mathfrak{f}_{k}^{\mathbb{I}}(n)
\hspace{0.1cm}+ \sum_{\mathclap{n=1}}^{\frac{12-p_{1}}{2}-1}\sum_{\mathclap{p_{1}=0,2,..}}^{10}\rho^{2n+p_{1}} \hspace{0.1cm}\mathfrak{f}_{p_{1}}^{\mathbb{II}}(n)
\hspace{0.1cm}+  \sum_{\mathclap{n=1}}^{\frac{13-p_{2}}{2}-1} & \sum_{\mathclap{p_{2}=1,3,..}}^{9}\rho^{2n+p_{2}} \hspace{0.1cm}\mathfrak{f}_{p_{2}}^{\mathbb{II}}(n)
\hspace{0.1cm}+\hspace{0.2cm}  \sum_{\mathclap{j=12,13,..}}^{\infty}\hspace{0.3cm}\left[\hspace{0.1cm}\sum_{\mathclap{k=0}}^{10} \hspace{0.1cm}\mathfrak{f}_{k}^{\mathbb{I}}(j-k) \hspace{0.3cm}+\hspace{0.3cm} \underbrace{\sum_{\mathclap{p=0,1,..}}^{10} \hspace{0.1cm}\mathfrak{f}_{p}^{\mathbb{II}}\left(\frac{j-p}{2} \right)\hspace{0.2cm}}_{\forall \hspace{0.05cm}(j-p)=0,2,4,...}\right]\rho^{j}
\end{align}
\end{footnotesize}
\be \label{eq:aa54}
= F_{0}\rho^{0}\hspace{0.1cm}+\hspace{0.1cm}F_{1}\rho^{1}\hspace{0.1cm}+\hspace{0.1cm}F_{2}\rho^{2}\hspace{0.1cm}+\hspace{0.1cm}...\hspace{0.1cm}+\hspace{0.1cm}F_{11}\rho^{11} \hspace{1cm}\left(\text{obviously,}\hspace{0.1cm}F_{0}=-\omega^2 r_{\text{\tiny 0}}^2,\hspace{0.2cm} F_{11}=0 \hspace{0.2cm}\text{from r.h.s of Eq}.\eqref{eq:A9}\right).
\ee
\end{widetext}
From the above Eq.\eqref{eq:aa54}, now one can evaluate the undetermined constants (\ti{viz.} $\Bbbk$, $\beta$, $a_{n}$-s) one by one which were introduced previously in Eqs.\eqref{eq:aa45} and \eqref{eq:aa46}. 
\begin{itemize}
\item
By equating the coefficients of $\rho^{0}$ on both sides of Eq.\eqref{eq:aa54}, we get
\be \label{eq:aa55}
-\Bbbk^{2}r_{\text{\tiny 0}}^{2} = -\omega^{2}r_{\text{\tiny 0}}^{2} \hspace{0.2cm}\Rightarrow \hspace{0.2cm} \Bbbk = \pm \omega.
\ee
Its to be noted that we take $\Bbbk=+\omega$ from now on in order to consider only the outgoing modes from the \textcolor{black}{sonic} horizon towards the external observer.

\item 
By equating the coefficients of $\rho^{1}$ on both sides of Eq.\eqref{eq:aa54}, we get 
\bea \nonumber
& 8 a_1 c_s \left(-r_{\text{\tiny 0}} \omega +2 i c_s\right)-4 c_s \Big(c_s (4 i
   \beta +\bar{l}+4)+r_{\text{\tiny 0}}^2 \mathcal{M} ^2 c_s \\ \nonumber 
& \hspace{6cm} -2 \beta  r_{\text{\tiny 0}} \omega \Big)=0,    \\ \nonumber
  & \Rightarrow \hspace{0.1cm}\beta = \frac{2 a_1 \left(4 c_s^2+r_{\text{\tiny 0}}^2 \omega ^2\right)+r_{\text{\tiny 0}} \omega  c_s
   \left(\bar{l} +\mathcal{M} ^2 r_{\text{\tiny 0}}^2+4\right)}{8 c_s^2+2 r_{\text{\tiny 0}}^2 \omega ^2} + i \Big(\frac{c_s^2 \left(\bar{l} +\mathcal{M} ^2 r_{\text{\tiny 0}}^2+4\right)}{4 c_s^2+r_{\text{\tiny 0}}^2 \omega
   ^2} \Big). \\ \label{eq:aa56}
\eea

\item 
By equating the coefficients of $\rho^{2}$ on both sides of Eq.\eqref{eq:aa54}, we get $a_{2}$ given by Eq.\eqref{eq:A21}, see Appendix-III.

\item
And so on and so forth, by equating the coefficients of $\rho^{11}$ on both sides of Eq.\eqref{eq:aa54} in the same manner, one can determine $a_{11}$ explicitly (in terms of $a_{1}$ which is kept nonzero arbitrary since the beginning). 

\item 
Finally, for any general $j$, one can find the coefficient $a_{j}$ ($\forall \hspace{0.05cm}j=12,13,14,...$) from the recursion relation which is deduced later in Appendix-III, see Eq.\eqref{eq:A20}.
\end{itemize}
It is to be noted that the recursion relation arises out of the square bracket on l.h.s of Eq.\eqref{eq:aa54} after having all the source terms fully exhausted. Though the explicit form of the mass term $\mathcal{M}$ is yet to be shown, the coefficients $\Bbbk$, $\beta$, $a_{n}$-s ($\forall \hspace{0.05cm}n \neq 1$) are all determined at this stage. Hence through the Eqs.\eqref{eq:aa45} and \eqref{eq:aa46}, one basically gets the full structure of the radial solution $\mathcal{R}(r)$.
\par 
In order to find $\mathcal{M}$ explicitly, one requires the exact form of $\vartheta_{1}(x)$ which is nothing but the solution of Eq.\eqref{eq:aa19}.

\subsection{4.3 \tu{Obtaining the usual mass-less scalar field $\vartheta_{1}(x)$}}
Like Eq.\eqref{eq:aa40}, one can consider an ansatz for the mass-less scalar field of the following form,
\bea \label{eq:aa58}
& \vartheta_{1}(x)=\frac{1}{r}\mathcal{R}_{1}(r)\mathcal{Y}^{l}_{\mathsf{m}}(\theta, \phi)e^{-i\omega_{1} \tau}, \\ \nonumber
& \text{typically,} \hspace{0.5cm} \omega_{1} >> \omega. \hspace{5cm}
\eea

Inserting this in Eq.\eqref{eq:aa38} with $\mathcal{M}=0$ gives rise to a radial equation (similarly like Eq.\eqref{eq:aa44}) of the following form,
\be \label{eq:aa77}
 \frac{d^{2} \mathcal{R}_{1}(r)}{d\chi^{2}}+\left[\omega_{1}^{2} - \left(\frac{\bar{l}}{r^{2}} +\frac{4r_{\text{\tiny 0}}^{4}}{r^{6}}\right)c_{s}^{2}\left(1-\frac{r_{\text{\tiny 0}}^{4}}{r^{4}}\right)\right] \mathcal{R}_{1}(r) =0. 
\ee 

Keeping the Eqs.\eqref{eq:aa45} and \eqref{eq:aa46} in mind, we pick $\mathcal{R}_{1}(r)$ to be of the following form,
\begin{align} \label{eq:aa59}
& \mathcal{R}_{1}(r)=\alpha_{1} e^{\pm i\hspace{0.05cm}\left[\Bbbk_{1} \hspace{0.05cm} \chi \hspace{0.1cm} +\hspace{0.1cm}  h_{1}(\rho) \right]}\hspace{0.05cm};& \\ \nonumber \mbox{where}\\ \nonumber
& \rho=1-\frac{r_{\text{\tiny 0}}}{r}, \hspace{0.3cm} h_{1}(\rho)=\beta_{1} \hspace{0.05cm}\ln(1-\rho)+\sum_{\mathclap{n=1}}^{\infty}\left(b_{n}\hspace{0.05cm}\rho^{n}\right),& \\ \label{eq:aa60}
& \hspace{4cm}\text{necessarily, $b_{1}\neq 0$}.&
\end{align}

By inspection, we figure out the following:
\begin{itemize}
\item
Exactly like Eq.\eqref{eq:aa55}, we conclude that 
\be \label{eq:aa61}
\Bbbk_{1}=\pm \omega_{1}
\ee
and we again take the `$+$' sign for the outgoing modes.

\item
If we just drop the mass term in Eq.\eqref{eq:aa56}, we simply come up with $\beta_{1}$. Therefore,
\bea \nonumber
 \beta_{1} & \equiv & \beta \vert_{\mathcal{M}=0} \hspace{6cm}\\ \label{eq:aa62}
&=& \left(  b_1+\frac{(\bar{l} +4) r_{\text{\tiny 0}} \omega_{1}  c_s}{8 c_s^2+2 r_{\text{\tiny 0}}^2 \omega_{1} ^2} \right) +i  \left( \frac{(\bar{l} +4) c_s^2}{4 c_s^2+r_{\text{\tiny 0}}^2 \omega _1^2}\right).
\eea
 
\item
Similarly, having the mass term dropped from the expression of $a_{2}$ in Eq.\eqref{eq:A21}, we come up with $b_{2}$ given by Eq.\eqref{eq:A22} in Appendix-III.

\item
And so on till $b_{11}$ in the same manner, followed by the recursion relation for some general $b_{j}$ ($\forall \hspace{0.05cm}j=12,13,14,...$) helps determine the rest of the coefficients explicitly and all in terms of $b_{1}$.
\end{itemize}

\subsubsection{4.3.1 \tb{Outside at a finite distance from the \textcolor{black}{sonic} horizon ($r\gtrsim r_{\text{\tiny 0}}$)}}

If one considers the mass-less solution (see Eq.\eqref{eq:aa58}) to be residing just outside the \textcolor{black}{sonic} horizon w.r.t some external observer, then the radial coordinate of $\vartheta_{1}(x)$ is obviously almost of the same order of $r_{\text{\tiny 0}}$, i.e., $r\gtrsim r_{\text{\tiny 0}}$.
\par 
In this regime, the measure of the variable $\rho=1-\frac{r_{\text{\tiny 0}}}{r}$ gives a very small number and thus one can fairly restrict oneself to the first order of $\rho$, while neglecting its higher powers throughout the calculations. And therefore, Eq.\eqref{eq:aa60} is approximated to 
\be \label{eq:aa64}
h_{1}(\rho) \approx \beta_{1} \hspace{0.05cm}\ln(1-\rho)+b_{1}\hspace{0.05cm}\rho.
\ee
Now inserting the Eqs.\eqref{eq:aa61}, \eqref{eq:aa62} and \eqref{eq:aa64} back into Eq.\eqref{eq:aa59}, we get
\begin{widetext}
\be \label{eq:aa65}
\mathcal{R}_{1}(r) \approx \alpha_{1} \exp\Big[\pm \hspace{0.1cm} i\omega_{1} \hspace{0.05cm} \chi \hspace{0.1cm}  \pm  \hspace{0.1cm}i b_{1}\left(1-\frac{r_{\text{\tiny 0}}}{r}\right)\hspace{0.1cm} \pm \hspace{0.1cm}\left\lbrace \frac{(\bar{l} +4) c_s^2}{4 c_s^2+r_{\text{\tiny 0}}^2 \omega _1^2} +i \left(-b_1-\frac{r_{\text{\tiny 0}} \omega_{1}  c_s (\bar{l} +4)}{8 c_s^2+2 r_{\text{\tiny 0}}^2 \omega_{1} ^2} \right) \right\rbrace \ln \frac{r}{r_{\text{\tiny 0}}}\Big].
\ee
And hence from Eq.\eqref{eq:aa58},
\be \label{eq:aa66}
\boxed{\vartheta_{1}(x)\approx \frac{1}{r}\hspace{0.1cm}\alpha_{1} \left(\frac{r}{r_{\text{\tiny 0}}}\right)^{\pm \frac{(\bar{l} +4) c_s^2}{4 c_s^2+r_{\text{\tiny 0}}^2 \omega _1^2}}\hspace{0.1cm}\exp\Bigg[\pm i \left\lbrace \omega_{1} \hspace{0.05cm} \chi \hspace{0.1cm}  +  \hspace{0.1cm} b_{1}\left(1-\frac{r_{\text{\tiny 0}}}{r}\right)\hspace{0.1cm} + \hspace{0.1cm} \left(-b_1-\frac{r_{\text{\tiny 0}} \omega_{1}  c_s (\bar{l} +4)}{8 c_s^2+2 r_{\text{\tiny 0}}^2 \omega_{1} ^2} \right)  \ln \frac{r}{r_{\text{\tiny 0}}}\right\rbrace\Bigg]\hspace{0.1cm}\mathcal{Y}^{l}_{\mathsf{m}}(\theta, \phi)e^{-i\omega_{1} \tau}}.
\ee
\end{widetext}
This above Eq.\eqref{eq:aa66} gives the final expression of the mass-less scalar field approximated upto linear $\rho$. Between the $\pm$ signs inside the exponent above, one should consider the `$-$' sign for the in-going modes from the sonic horizon towards the center of the acoustic black hole, and the `$+$' sign for the out-going modes. An important thing to be noted here is that the spatial growth of these short wavelength modes goes as $\sim \hspace{0.02cm} r^{\left(\omega_{1}^{-2}\right)}$. This would be later compared with the growth of the large wavelength amplitude modes $\varphi(x)$.
\par 
Now we move ahead to find the expression of the mass term.

\subsection{4.4 \tu{Deriving the full mass term $\mathcal{M}$}}

We can identify the following steps we had taken towards arriving at the complete expression of $\mathfrak{m}$, for a canonical acoustic BH, given by Eq.\eqref{eq:A1} (see Appendix-I):
\begin{enumerate}

\item 
Eq.\eqref{eq:aa18} was applied on Eq.\eqref{eq:aa17} throughout keeping the small-scale Laplacians and the background velocity field \tb{v}$(x)$ unperturbed. Thus we ended up with an enormously large set of terms and got them all segregated according to the different orders of the parameter $\epsilon$.
\item 
Out of that huge lot, we collected a pack of terms at $\mathcal{O}(\epsilon^{2})$ and equated their sum total to zero in order to form an equation in larger scales where $\vartheta_{1}(x)$ was treated effectively as a constant.  

\item 
Among the terms written on the l.h.s of this equation, a number of terms got compacted as \textcolor{black}{$\partial_{\bar{\mu}}f_{2}^{\bar{\mu} \bar{\nu}} \partial_{\bar{\nu}} \vartheta_{2}(X)$} while the rest were being identified as something proportional to the amplitude field, i.e. $-\mathfrak{m}^{2}\vartheta_{2}(X)$. The expression of $\mathfrak{m}$, until this step, would naturally contain a factor of $R^{2}\sin\Theta$ while appearing in Eq.\eqref{eq:aa21}.

\item 
After the scale-reversion (see Section 3.2), this factor of \hspace{0.05cm}$R^{2}\sin\Theta$\hspace{0.05cm} simply became \hspace{0.05cm} $r^{2}\sin\theta$\hspace{0.05cm} and gave rise to the expression of a re-scaled mass $\mathfrak{m}$ inserted in Eq.\eqref{eq:aa25}. 

\item 
In order to keep things from getting too messy and unnecessarily cluttered, we consider the background velocity to be $|\tb{v}(\tb{r})|=v_{r}(r)$; which is exactly nothing but the case for a canonical acoustic black hole, see Eq.\eqref{eq:aa33}. Thus we get a bit tidier expression for $\mathfrak{m}$, given by Eq.\eqref{eq:A1}.

\end{enumerate}

With $\vartheta_{1}(x)$ field in hand, as shown\footnote{Out of two possible signs, we are interested only in the outgoing modes and hence consider the $+i$ in the exponent of the r.h.s of Eq.\eqref{eq:aa66} throughout the paper.} in Eq.\eqref{eq:aa66}, one can just readily evaluate $\mathfrak{m}\vert_{\text{\tiny{for Canonical BH}}}$ through Eq.\eqref{eq:A1}. Now finding the final mass term $\mathcal{M}$ for a canonical acoustic black hole is simply redundant and a one-step-process. Using Eq.\eqref{eq:aa31}, 
\be  \label{eq:aa68}
\mathcal{M}^{2}\vert_{\text{\tiny{for Canonical BH}}}=\frac{1}{c_{s}r^{2}\sin\theta}\frac{\mathsf{g}}{c_{s}}\epsilon^{2}\hspace{0.1cm}\mathfrak{m}^{2}\vert_{\text{\tiny{for Canonical BH}}}.
\ee
Since our formalism is restricted only within the domain of a canonical acoustic black hole, from now on, we would call off the subscript for the mass term(s) and write just $\mathcal{M}$ to refer to the mass term as expressed by the above Eq.\eqref{eq:aa68}. 



While deriving $\mathcal{M}$, we would again restrict ourselves to considering only the most dominant term(s).

After some trivial and tedious algebra, the expression of the mass term is finally given by,
\begin{widetext}
\bea \label{eq:3new}
 \mathcal{M} \hspace{0.1cm}&=&\hspace{0.1cm}  \xi \bigg[\left(\frac{-176c_{s}^{2}\omega_{1}^{2}+r_{\text{\tiny 0}}^2 \omega_{1}^{4}}{256c_{s}^{4}r_{\text{\tiny 0}}^{2}} + i \hspace{0.1cm}\frac{-48c_{s}^{2}\omega_{1}+3r_{\text{\tiny 0}}^2 \omega_{1}^{3}}{32c_{s}^{3}r_{\text{\tiny 0}}^{3}}\right)\rho^{-4} \hspace{0.1cm}+\hspace{0.1cm} \mathcal{O}(\rho^{-3})\hspace{0.1cm}+\hspace{0.1cm}... \bigg]^{1/2}\hspace{0.5cm} ,\\ \nonumber \\ \label{eq:4new}
&\approx &\hspace{0.1cm} \mathcal{M}_{\mathcal{O}(\rho^{-4})} \hspace{0.5cm},
\eea
\bea \nonumber 
& \text{where,} \hspace{14cm}\\ \label{eq:aa70}
& \mathcal{M}_{\mathcal{O}(\rho^{-4})}^{2} = \boxed{\frac{\xi^{2}}{\left(1-\frac{r_{\text{\tiny 0}}}{r}\right)^{4}}\frac{1}{32 c_{s}^{4}\hspace{0.05cm}r_{\text{\tiny 0}}^{3}}\bigg(\frac{r_{\text{\tiny 0}}}{8}\left(-176c_{s}^{2}\omega_{1}^{2}+r_{\text{\tiny 0}}^2 \omega_{1}^{4}\right) + i \hspace{0.1cm}3c_{s}\left(-16c_{s}^{2}\omega_{1}+r_{\text{\tiny 0}}^2 \omega_{1}^{3}\right)\bigg)}.
\eea
\end{widetext}
It is interesting to note that, as far as the most dominant terms are concerned, $\mathcal{M}_{\mathcal{O}(\rho^{-4})}$ happens to be independent of the choice of a particular spherical harmonic while picking $\vartheta_{1}(x)$ from Eq.\eqref{eq:aa66}. Hence this is the most general expression of the mass term of the phonon modes associated to $\varphi(x)$ field in a canonical spacetime within the regime not too far from the sonic horizon. 
\par 
One may be interested in a more accurate measure of $\mathcal{M}$, and hence the sub-leading contributions could be relevant in that scenario. See Eq.\eqref{eq:A23} in Appendix-I, where we have given the next two sub-leading contributions in the expression of the mass term.
\par 
As it is quite evident from the above Eq.\eqref{eq:aa70} that $\mathcal{M}_{\mathcal{O}(\rho^{-4})}$ does depend on the position $r$ and this kind of coordinate dependence of the mass term appears in many contexts of physics - e.g.\cite{Visser2005} where Visser \ti{et al.} encountered a position-dependent-mass. But if $\mathcal{M}$ is picked from Eq.\eqref{eq:3new}, then for an observer sitting at a very large $r$, the mass term becomes a real constant, i.e. $\lim_{r\to\infty}\mathcal{M}=\frac{\xi \omega_{1}^{2}}{c_{s}^{2}}$ in the asymptotic limit for any arbitrary $\omega_{1}\in \Re$.
\par 
To keep the notations simpler, from now on, we would refer to $\mathcal{M}_{\mathcal{O}(\rho^{-4})}$ in Eq.\eqref{eq:aa70} by calling it just as $\mathcal{M}$ since the following calculations are solely based only on the leading order contributions in the mass term.

\subsection{4.5 \tu{Obtaining the \ti{massive} scalar field $\varphi(x)$}}

Finally we are on the verge of deriving the expression of massive scalar field. Like Eq.\eqref{eq:aa65}, one can now obtain the radial contribution to the massive field upto linear $\rho$ in order to consider only the leading order contributions. 
\par
With the mass term $\mathcal{M}$ in hand, one obtains $\beta$ from Eq.\eqref{eq:aa56} and then using Eqs.\eqref{eq:aa45} and \eqref{eq:aa46}, we come up with
\begin{widetext}
\be \label{eq:aa71}
\mathcal{R}(r) \approx \alpha \exp\Big[\pm \hspace{0.1cm} i\omega \hspace{0.05cm} \chi \hspace{0.1cm}  \pm  \hspace{0.1cm}i a_{1}\left(1-\frac{r_{\text{\tiny 0}}}{r}\right)\hspace{0.1cm} \pm \hspace{0.1cm}\left\lbrace  \left(\frac{c_s^2 \left(\bar{l} +\mathcal{M} ^2 r_{\text{\tiny 0}}^2+4\right)}{4 c_s^2+r_{\text{\tiny 0}}^2 \omega
   ^2} \right) + i\left(-a_{1}-\frac{r_{\text{\tiny 0}} \omega  c_s
   \left(\bar{l} +\mathcal{M} ^2 r_{\text{\tiny 0}}^2+4\right)}{8 c_s^2+2 r_{\text{\tiny 0}}^2 \omega ^2} \right)\right\rbrace \ln \frac{r}{r_{\text{\tiny 0}}}\Big].
\ee
Thus the massive scalar field (see Eq.\eqref{eq:aa40}) is finally given by,

\ben 
 \varphi(x)\approx \frac{1}{r}\hspace{0.1cm}\alpha \exp\Bigg[\pm \hspace{0.1cm} i\omega \hspace{0.05cm} \chi \hspace{0.1cm}  \pm  \hspace{0.1cm}i a_{1}\left(1-\frac{r_{\text{\tiny 0}}}{r}\right)\hspace{0.1cm} \pm \hspace{0.1cm}\left\lbrace  \left(\frac{c_s^2 \left(\bar{l} +\mathcal{M} ^2 r_{\text{\tiny 0}}^2+4\right)}{4 c_s^2+r_{\text{\tiny 0}}^2 \omega
   ^2} \right) + i\left(-a_{1}-\frac{r_{\text{\tiny 0}} \omega  c_s
   \left(\bar{l} +\mathcal{M} ^2 r_{\text{\tiny 0}}^2+4\right)}{8 c_s^2+2 r_{\text{\tiny 0}}^2 \omega ^2} \right)\right\rbrace \ln \frac{r}{r_{\text{\tiny 0}}}\Bigg]\hspace{0.1cm}\mathcal{Y}^{l}_{\mathsf{m}}(\theta, \phi)e^{-i\omega \tau},
 \een
\begin{small}
\bea \nonumber
&= \frac{1}{r}\alpha \left(\frac{r}{r_{\text{\tiny 0}}}\right)^{\pm \left(\frac{2c_{s}^{2}\left(\bar{l}+4\right)+2c_{s}^{2}r_{\text{\tiny 0}}^{2}\Re(\mathcal{M}^{2})+r_{\text{\tiny 0}}^{3}\omega \Im(\mathcal{M}^{2})}{8 c_s^2+2 r_{\text{\tiny 0}}^2 \omega ^2} \right)}\exp\Bigg[\pm  i \left\lbrace \omega  \chi  + a_{1}\left(1-\frac{r_{\text{\tiny 0}}}{r}\right) - \left(a_{1}+\frac{c_{s}r_{\text{\tiny 0}}\Big(\omega \left(\bar{l}+4\right)+r_{\text{\tiny 0}}^{2}\omega \Re(\mathcal{M}^{2})-2c_{s}r_{\text{\tiny 0}}\Im(\mathcal{M}^{2})\Big)}{8 c_s^2+2 r_{\text{\tiny 0}}^2 \omega ^2}\right)\ln \frac{r}{r_{\text{\tiny 0}}}\right\rbrace \Bigg] \\ \nonumber 
&\hspace{16.5cm}\times \hspace{0.1cm}\mathcal{Y}^{l}_{\mathsf{m}}(\theta, \phi)e^{-i\omega \tau},
 \\ \label{eq:aa72}
\eea
\end{small}
\end{widetext}
where $\mathcal{M}$ is picked from Eq.\eqref{eq:aa70} with $\Re{\left(\mathcal{M}^{2}\right)}$ and $\Im{\left(\mathcal{M}^{2}\right)}$ being its real and imaginary parts respectively.
\par 
The above expression clearly indicates that at a fixed $r$, the growth rate over space is $\sim \hspace{0.02cm} r^{\frac{\|\mathcal{M}^{2}\|}{\omega^{2}}}$. So, from Eq.\eqref{eq:aa70}, the growth rate of these large wavelength modes, for a specific mode of frequency $\omega$, turns out to be actually $\sim \hspace{0.02cm} r^{\left(\omega_{1}^{4} \right)}$ which encodes the information of the supposedly Hawking radiated modes (i.e. $\vartheta_{1}(x)$ field). This gives rise to the low frequency (or larger wavelength) band of $\omega$, i.e. $\varphi(x)$ field which, in the absence of the mass term (or in other words when $\xi^{2}\approx 0$), is not distinguishable from the primary $\omega_{1}$-modes. Obviously the smaller $\omega$-modes would grow faster and effectively extract more energy from the $\omega_{1}$-modes which are supposedly Hakwing radiated.

\section{5. Discussion}

In the present paper, we systematically analyzed the consequences of the presence of the quantum potential term in the dynamics of a condensate on the perspectives of analogue Hawking radiation. Here we have worked out this formulation for a canonical acoustic BH configuration in $(3+1)$-d spacetime. The quantum potential term causes a UV-IR coupling which can be separated as an independent dynamics at larger length scales without disturbing the Lorentz invariance of the basic KG equation (massless) is something that we have already shown \cite{Sarkar2016} and the present work extends the same method to curved spacetime.
\par 
The presence of the UV-IR coupling resulting from the quantum potential would make short wavelength modes to lose energy to large wavelength ones which show up as massive amplitude excitations of the high frequency Hawking radiated modes. In the actual experimental evaluation of analogue Hawking radiation, one can not neglect these large wavelength modes which will grow from primary Hawking radiated quanta and would cause an "information loss" of the actually Hawking radiated modes.
\par 
Our present analysis shows that the growth rate of these large wavelength ($\omega$) modes, in a canonical spacetime, holds the clue to keep the underlying physics consistent. In general, a massless scalar field would grow over space near the analogue \textcolor{black}{acoustic Killing / sonic} horizon (the region which is accessible in the experiments) as something $\sim \hspace{0.02cm} r^{\left(\omega_{1}^{-2}\right)}$. On the contrary, the massive secondary excitations generated by these primary modes would grow over space as $\sim \hspace{0.02cm} r^{\left(\omega_{1}^{4}\hspace{0.03cm}\omega^{-2}\right)}$ for large $\omega_{1}$, but $\omega$ can be obtained easily from the temporal profile, i.e. $e^{-i\omega \tau}$ in Eq.\eqref{eq:aa72}, of the large wavelength signal as received by the external observer. So a careful observation of the $\omega_{1}$ dependance of the growth rates of these secondary modes can actually reveal the relative abundance of the originally Hawking radiated quanta in the $(3+1)$-d canonical spacetime. 
\par 
These massive amplitude modes arise from the quantum connection which is $\mathcal{O}(\epsilon^{2})$ small. But, at the same time, one should be insured that $\omega_{1}$ is typically large and that makes this mechanism of secondary excitation generation absolutely relevant in the quantum fluids, like BEC. 
\par 
We present in this paper a detailed derivation and analysis of these excitations generated by quantum potential which, in every likelihood, would be a dominant contributor to the loss of correlations which are instrumented in probing the analogue Hawking effect in such systems. We hope to extend our present analysis in deriving the correction to the correlations of Hawking radiated quanta in other low-dimensional experimentally relevant systems within the scope of our framework.

\section*{ACKNOWLEDGMENTS}
SS would like to acknowledge fruitful discussions with \tb{Shankhadeep Chakrabortty}.


\section*{Appendices}

In order to tackle some untidy and too cluttered expressions in a proper presentable manner, we will be introducing a few new symbols here, viz. $\mathcal{Q}_{1}$, $\mathcal{Q}_{2}$, etc., whenever required.

\subsection*{\tu{Appendix-I}}
The following expression of $\mathfrak{m}$ is written in case of a canonical acoustic black hole, where the background velocity is picked by Eq.\eqref{eq:aa33}.
\begin{widetext}
\begin{small}
\begin{subequations} \label{eq:A1}
\bea \label{eq:A1A}
& \mathfrak{m}^{2}\vert_{\text{\tiny{for Canonical BH}}}=\frac{\xi_{0}^{2}}{\mathsf{g}} \hspace{0.05cm}\frac{\sin\theta}{r^{2}} \hspace{0.05cm}\frac{1}{\vartheta_{1}(x)} \hspace{0.05cm} \mathcal{Q}_{1}(x)\vert_{\text{\tiny{for Canonical BH}}}, \hspace{1cm}\text{where,} \hspace{8cm}\\ \nonumber 
& \mathcal{Q}_{1}(x)\vert_{\text{\tiny{for Canonical BH}}} \equiv \mathcal{Q}_{1}(x)\vert_{v_{\theta}=0=v_{\phi}} \hspace{13cm}\\ \nonumber 
&\hspace{2cm} = \Bigg[ r^{4}v_{r}^{\prime \prime}\frac{\partial^{2}\vartheta_{1}}{\partial t \hspace{0.05cm}\partial r} 
+ 2r^{3}\frac{\partial^{3}\vartheta_{1}}{\partial t^{2} \hspace{0.05cm}\partial r}
+ 4r^{3} v_{r}^{\prime}\frac{\partial^{2}\vartheta_{1}}{\partial t \hspace{0.05cm}\partial r}
+ r^{2} \frac{\partial^{4}\vartheta_{1}}{\partial t^{2} \hspace{0.05cm} \partial \theta^{2}}
+ \frac{r^{2}}{\sin^{2}\theta}\frac{\partial^{4}\vartheta_{1}}{\partial t^{2} \hspace{0.05cm} \partial \phi^{2}}
+ r^{2}\cot\theta \frac{\partial^{3}\vartheta_{1}}{\partial t^{2} \hspace{0.05cm} \partial \theta}
+ r^{2} v_{r}^{\prime}\frac{\partial^{3}\vartheta_{1}}{\partial t \hspace{0.05cm} \partial \theta^{2}}
+ \frac{r^{2}}{\sin^{2}\theta} v_{r}^{\prime}\frac{\partial^{3}\vartheta_{1}}{\partial t \hspace{0.05cm} \partial \phi^{2}} \\ \nonumber
&\hspace{1.2cm}+ r^{2}\cot\theta \hspace{0.05cm}v_{r}^{\prime}\frac{\partial^{2}\vartheta_{1}}{\partial t  \hspace{0.05cm} \partial \theta}
+ r^{4} \frac{\partial^{4}\vartheta_{1}}{\partial t^{2} \hspace{0.05cm} \partial r^{2} }
+ 3r^{4} v_{r}^{\prime}\frac{\partial^{3}\vartheta_{1}}{\partial t \hspace{0.05cm} \partial r^{2}} 
+ v_{r}\bigg\lbrace 
2r^{2}\frac{\partial^{4}\vartheta_{1}}{\partial t \hspace{0.05cm} \partial r \hspace{0.05cm} \partial \theta^{2}}
+ \frac{2r^{2}}{\sin ^{2}\theta}\frac{\partial^{4}\vartheta_{1}}{\partial t \hspace{0.05cm} \partial r \hspace{0.05cm} \partial \phi^{2}}
+ 2r^{2}\cot\theta \frac{\partial^{3}\vartheta_{1}}{\partial t \hspace{0.05cm} \partial r \hspace{0.05cm} \partial \theta}
+ 2r^{2}\frac{\partial^{2}\vartheta_{1}}{\partial t \hspace{0.05cm} \partial r } \\ \nonumber
& \hspace{1.2cm}+ 2r^{2}v_{r}^{\prime}\frac{\partial^{3}\vartheta_{1}}{\partial r \hspace{0.05cm} \partial \theta^{2}}
+ \frac{2r^{2}}{\sin ^{2}\theta} v_{r}^{\prime}\frac{\partial^{3}\vartheta_{1}}{ \partial r \hspace{0.05cm} \partial \phi^{2}}
+ 2r^{2}\cot\theta \hspace{0.05cm} v_{r}^{\prime}\frac{\partial^{2}\vartheta_{1}}{\partial r \hspace{0.05cm} \partial \theta}
+ 2r^{4}\frac{\partial^{4}\vartheta_{1}}{\partial t \hspace{0.05cm} \partial r^{3} }
+ 6r^{3}\frac{\partial^{3}\vartheta_{1}}{\partial t \hspace{0.05cm} \partial r^{2} }
+ r^{4}v_{r}^{(3)}\frac{\partial \vartheta_{1}}{\partial r}
+ 4r^{3}v_{r}^{\prime \prime}\frac{\partial \vartheta_{1}}{ \partial r }
+ 2r^{2}v_{r}^{\prime}\frac{\partial \vartheta_{1}}{\partial r} \\ \nonumber
& \hspace{1.2cm} + 4r^{4}v_{r}^{\prime}\frac{\partial^{3}\vartheta_{1}}{\partial r^{3} }
+ 3r^{4}v_{r}^{\prime \prime}\frac{\partial^{2}\vartheta_{1}}{ \partial r^{2}}
+ 10r^{3}v_{r}^{\prime}\frac{\partial^{2}\vartheta_{1}}{ \partial r^{2} }
\bigg\rbrace
+ r^{2}v_{r}^{2}\bigg( 
\frac{\partial^{4}\vartheta_{1}}{\partial r^{2} \hspace{0.05cm} \partial \theta^{2} } 
+ \frac{1}{\sin ^{2}\theta} \frac{\partial^{4}\vartheta_{1}}{\partial r^{2} \hspace{0.05cm} \partial \phi^{2} } 
+ \cot\theta \frac{\partial^{3}\vartheta_{1}}{\partial r^{2} \hspace{0.05cm} \partial \theta } 
+ 4r \frac{\partial^{3}\vartheta_{1}}{\partial r^{3} } 
+ 2\frac{\partial^{2}\vartheta_{1}}{\partial r^{2} } 
+ r^{2}\frac{\partial^{4}\vartheta_{1}}{ \partial r^{4} } 
\bigg) \\ \label{eq:A1B}
& + r^{4}v_{r}^{\prime}v_{r}^{\prime \prime} \frac{\partial \vartheta_{1}}{\partial r} 
+  2r^{3}\left(v_{r}^{\prime}\right)^{2} \frac{\partial \vartheta_{1}}{\partial r}
+ 2r^{4}\left(v_{r}^{\prime}\right)^{2} \frac{\partial^{2} \vartheta_{1}}{\partial r^{2}}
\Bigg].
\eea
\end{subequations}
\end{small}
\end{widetext}
In the above expression, $v_{r}^{\prime}\equiv \frac{\partial v_{r}}{\partial r}$, $v_{r}^{\prime \prime}\equiv \frac{\partial^{2} v_{r}}{\partial r^{2}}$, and $v_{r}^{(3)}\equiv \frac{\partial^{3} v_{r}}{\partial r^{3}}$.
\par 
Only upto leading order contributions are considered in the final expression of the mass term in Eq.\eqref{eq:aa70}. Here we have given a more accurate measure by taking into account the next two sub-leading contributions, given by,
\be \label{eq:A23}
 \mathcal{M} \approx \Big[\mathcal{Q}_{3}(r) \hspace{0.1cm}+\hspace{0.1cm} i \mathcal{Q}_{4}(r)\Big]^{1/2}, 
\ee
\begin{subequations} \label{eq:A24}
\bea \nonumber
& \text{where,} \hspace{11cm}\\ \nonumber 
& \mathcal{Q}_{3}(r)= \frac{\xi^{2}\omega_{1}^{2}\left(-176c_{s}^{2}+r_{\text{\tiny 0}}^2 \omega_{1}^{2}\right)}{256c_{s}^{4}r_{\text{\tiny 0}}^{2}\left(1-\frac{r_{\text{\tiny 0}}}{r}\right)^{4}} 
+ \frac{\xi^{2}\omega_{1}^{2}\big(64c_{s}^{4}( 88+7\bar{l} ) -8c_{s}^{2}(-113+\bar{l} )r_{\text{\tiny 0}}^{2}\omega_{1}^{2} - 6r_{\text{\tiny 0}}^{4} \omega_{1}^{4}\big)}{256c_{s}^{4}r_{\text{\tiny 0}}^{2}\left(4c_{s}^{2}+r_{\text{\tiny 0}}^{2}\omega_{1}^{2}\right)\left(1-\frac{r_{\text{\tiny 0}}}{r}\right)^{3}} \\ \nonumber
&+ \frac{\xi^{2}\omega_{1}^{2}\bar{l}}{16c_{s}^{2}r_{\text{\tiny 0}}^{2}\left(1-\frac{r_{\text{\tiny 0}}}{r}\right)^{2}} \hspace{0.5cm}, 
\eea
\\ \\ and 
\bea \nonumber
& \mathcal{Q}_{4}(r)= \frac{\xi^{2}\omega_{1}\left(-48c_{s}^{2}+3r_{\text{\tiny 0}}^2 \omega_{1}^{2}\right)}{32c_{s}^{3}r_{\text{\tiny 0}}^{3}\left(1-\frac{r_{\text{\tiny 0}}}{r}\right)^{4}} 
+ \frac{\xi^{2}\omega_{1}\big(64c_{s}^{4}( 20+ \bar{l} ) +2c_{s}^{2}(66-7\bar{l} )r_{\text{\tiny 0}}^{2}\omega_{1}^{2} - 17r_{\text{\tiny 0}}^{4} \omega_{1}^{4}\big)}{32c_{s}^{3}r_{\text{\tiny 0}}^{3}\left(4c_{s}^{2}+r_{\text{\tiny 0}}^{2}\omega_{1}^{2}\right)\left(1-\frac{r_{\text{\tiny 0}}}{r}\right)^{3}} \\ \nonumber
&+ \frac{\xi^{2}\omega_{1}\bar{l}}{4c_{s}r_{\text{\tiny 0}}^{3}\left(1-\frac{r_{\text{\tiny 0}}}{r}\right)^{2}} \hspace{0.5cm}. 
\eea
\end{subequations}


\vspace{2cm}
\subsection*{\tu{Appendix-II}}

In this sub-section, we would investigate the nature of singularities of Eq.\eqref{eq:aa41} at various points. We first rewrite it in the following manner
\begin{widetext}
\be \label{eq:A2}
\frac{d^{2}}{dr^{2}}\mathcal{R}(r)+\underbracket{\frac{4 r_{\text{\tiny 0}}^{4}}{r \left(r^{4}-r_{\text{\tiny 0}}^{4}\right)}}_{\substack{\text{$\mathsf{p}(r)$\hspace{0.2cm}(say)}}}\frac{d}{dr}\mathcal{R}(r)+\underbracket{\frac{ \omega^{2}r^{10}-c_{s}^{2} \mathcal{M}^{2} \left(r^{4}-r_{\text{\tiny 0}}^{4}\right) r^{6}-c_{s}^{2} \left(r^{4}-r_{\text{\tiny 0}}^{4}\right) \Big(l(l+1) r^{4} +4 r_{\text{\tiny 0}}^{4}\Big)}{c_{s}^{2} r^{2} \left(r^{4}-r_{\text{\tiny 0}}^{4}\right){}^2}}_{\substack{\text{$\mathsf{q}(r)$\hspace{0.2cm}(say)}}}\mathcal{R}(r) = 0.
\ee 
\end{widetext}
\begin{itemize}
\item
At $r=0$ \hspace{0.05cm}: \hspace{0.05cm} $ \lim_{r \to 0} \mathsf{p}(r) \rightarrow -\infty $, \hspace{0.02cm} $ \lim_{r \to 0} \mathsf{q}(r) \rightarrow \infty , $ 
i.e. There's a singularity at $r=0$ point. But we find that 
\begin{align} \label{eq:A3}
&\lim_{r \to 0} \big[r\mathsf{p}(r)\big] =-4,  &\lim_{r \to 0} \big[r^{2}\mathsf{q}(r)\big] =4,  
\end{align}
which refer to $r=0$ to be a \ti{regular} singular point.
\item
At $r=r_{\text{\tiny 0}}$ \hspace{0.05cm}:\hspace{0.05cm} $ \lim_{r \to r_{\text{\tiny 0}}} \mathsf{p}(r) \rightarrow \infty $, \hspace{0.02cm} $ \lim_{r \to r_{\text{\tiny 0}}} \mathsf{q}(r) \rightarrow \infty , $ 
i.e. There's again a singularity at $r=r_{\text{\tiny 0}}$ point. But
\begin{align} \nonumber
&\lim_{r \to r_{\text{\tiny 0}}} \Big[(r-r_{\text{\tiny 0}})\mathsf{p}(r)\Big] =1,  &\lim_{r \to r_{\text{\tiny 0}}}  \Big[(r-r_{\text{\tiny 0}})^{2}\mathsf{q}(r)\Big] =\frac{\omega^{2}r_{\text{\tiny 0}}^{2}}{16c_{s}^{2}},  \\ \label{eq:A4}
\end{align}
hold true which guarantee that $r=r_{\text{\tiny 0}}$ is a \ti{regular} singular point.
\item
At $r\to \infty$ \hspace{0.05cm}: \hspace{0.05cm}We take $r=\frac{1}{r_{*}}$, thus as $r\to \infty$, $r_{*} \to 0$. Now obviously $ \mathsf{p}(r=\frac{1}{r_{*}}) $ becomes some $\mathsf{p}_{1}(r_{*})$ and $ \mathsf{q}(r=\frac{1}{r_{*}}) $ becomes some $\mathsf{q}_{1}(r_{*})$. We have,
\begin{align} \label{eq:A5}
&\lim_{r_{*} \to 0} \Big[\frac{2}{r_{*}}-\frac{\mathsf{p}_{1}(r_{*})}{r_{*}^{2}}\Big] \rightarrow \infty,  &\lim_{r_{*} \to 0} \Big[\frac{\mathsf{q}_{1}(r_{*})}{r_{*}^{4}}\Big] \rightarrow \infty,  
\end{align}
which basically mean $r_{*} \to 0$ or $r \to \infty$ is a singular point; while 
\begin{align} \label{eq:A6}
&\lim_{r_{*} \to 0} \Big[r_{*}\left(\frac{2}{r_{*}}-\frac{\mathsf{p}_{1}(r_{*})}{r_{*}^{2}}\right)\Big] =2,  &\underbrace{\lim_{r_{*} \to 0} \Big[r_{*}^{2}\left(\frac{\mathsf{q}_{1}(r_{*})}{r_{*}^{4}}\right)\Big] \rightarrow \infty}_{\text{i.e. the limit doesn't exist}}. 
\end{align}
\end{itemize}

\subsection*{\tu{Appendix-III}}

Inserting Eqs.\eqref{eq:aa45} and \eqref{eq:aa46} into the Eq.\eqref{eq:aa44}, we calculate the l.h.s of Eq.\eqref{eq:aa44} part by part.
\begin{widetext}
\bea \nonumber
& \text{The first part of l.h.s of Eq.\eqref{eq:aa44}} \hspace{13cm}
\eea
\bea \nonumber
 \Rightarrow \frac{d^{2} \mathcal{R}(r)}{d\chi^{2}} &=& \mathcal{Q}_{2}(r) \times \hspace{0.05cm}
\left(-\frac{1}{r^{10}} \right)\Bigg[ \Bbbk^{2}r^{10}+ \beta c_{s}^{2}\left(r^{4}-r_{\text{\tiny 0}}^{4}\right)\left(r^{4}(-i+\beta)-(-5i+\beta)r_{\text{\tiny 0}}^{4}\right)- 2\Bbbk r^{5}\beta c_{s} \left(r^{4}-r_{\text{\tiny 0}}^{4}\right)  \\ \nonumber
&+& \sum_{\mathclap{n=1}}^{\infty} \bigg\lbrace - i n a_{n}c_{s} r_{\text{\tiny 0}}\underbrace{\left(1-\frac{r_{\text{\tiny 0}}}{r}\right)^{n}}\left(r^3+r_{\text{\tiny 0}} r^2+r_{\text{\tiny 0}}^2 r+r_{\text{\tiny 0}}^3\right) \bigg(c_s r_{\text{\tiny 0}} r^3 (n -1)+c_s r_{\text{\tiny 0}}^2 r^2
   (n -1)+c_s r_{\text{\tiny 0}}^4 (2 i \beta +n +5) \\ \label{eq:A7}
&+& c_s r_{\text{\tiny 0}}^3 r (n -1)  +c_s (-2-2 i \beta )r^4  +2 i \Bbbk r^5\bigg) + n^2 a_n^2 c_s^2 r_{\text{\tiny 0}}^2 \underbrace{\left(1-\frac{r_{\text{\tiny 0}}}{r}\right)^{2 n}} \left(r^3+r_{\text{\tiny 0}} r^2+r_{\text{\tiny 0}}^2 r+r_{\text{\tiny 0}}^3\right)^2 \bigg \rbrace
\Bigg],
\eea
\end{widetext}

where the above two terms are under-braced for a reason because, while switching from $r \rightarrow \rho$, these two terms obviously become $\rho^{n}$ and $\rho^{2n}$ respectively which are boxed in Eq.\eqref{eq:aa51}.
\begin{widetext}
\bea \nonumber
& \text{And the second part of l.h.s of Eq.\eqref{eq:aa44}} \hspace{13cm}
\eea
\bea \label{eq:A8}
& \Rightarrow \left[\omega^{2} - \left(\mathcal{M}^{2}+\frac{\bar{l}}{r^{2}} +\frac{4r_{\text{\tiny 0}}^{4}}{r^{6}}\right)c_{s}^{2}\left(1-\frac{r_{\text{\tiny 0}}^{4}}{r^{4}}\right)\right] \mathcal{R}(r) \hspace{0.2cm}=\hspace{0.2cm}
\mathcal{Q}_{2}(r) \times \Bigg[ \omega ^2-\frac{ c_s^2 \left(r^4-r_{\text{\tiny 0}}^4\right)\left(\bar{l}\hspace{0.05cm} r^4+\mathcal{M}^2 r^6+4 r_{\text{\tiny 0}}^4\right)}{r^{10}} \Bigg],  \\ \nonumber \\ \nonumber \\ \nonumber
& \text{where,} \hspace{0.5cm} \mathcal{Q}_{2}(r)= \alpha \hspace{0.05cm} \exp \left[i \hspace{0.05cm}\left\lbrace \frac{ \Bbbk }{4 c_s}\left(4r -2r_{\text{\tiny 0}}\tan ^{-1}\left(\frac{r}{r_{\text{\tiny 0}}}\right) +r_{\text{\tiny 0}}\ln \left|\frac{1-\frac{r}{r_{\text{\tiny 0}}}}{1+\frac{r}{r_{\text{\tiny 0}}}}\right|\hspace{0.1cm}\right)+ \sum_{n} \left(a_n \left(1-\frac{r_{\text{\tiny 0}}}{r}\right) ^{n} \right) \right\rbrace \right] \hspace{0.05cm} \left(\frac{r_{\text{\tiny 0}}}{r}\right)^{i\beta}. 
\eea
\end{widetext}

Clearly, $\mathcal{Q}_{2}(r) \neq 0$ because $\alpha \neq 0$. Having the above two Eqs.\eqref{eq:A7} and \eqref{eq:A8} attached together, followed by a trivial algebraic manipulation, we actually come up with Eq.\eqref{eq:aa51}. Now we present the corresponding coefficients of $\rho^{n}$, $\rho^{n+1}$,... $\forall \hspace{0.05cm}n=1(1)\infty$ in a more jotted down way by rewriting the parent equation (i.e. Eq.\eqref{eq:aa44}) in the following manner,
\begin{widetext}
\begin{footnotesize}
\begin{align} \nonumber
& \underbrace{-\Bbbk^2 r_{\text{\tiny 0}}^2 }\hspace{0.1cm}+\hspace{0.1cm}\sum_{\mathclap{n=1}}^{\infty}\hspace{0.05cm}\Bigg[\hspace{0.1cm}8 i n a_{n} c_s \left(2 n c_s+i \Bbbk r_{\text{\tiny 0}}\right)\underbrace{\rho^{n}}+4 n a_{n} c_s \left(7 \Bbbk r_{\text{\tiny 0}}+2 c_s (4 \beta -14 i n-7 i)\right)\underbrace{\rho ^{n+1}}+4 i n a_{n} c_s  \left(c_s (48 i \beta +89
   n+89)+10 i \Bbbk r_{\text{\tiny 0}}\right) \underbrace{\rho ^{n+2}} \\ \nonumber   
& \hspace{0.5cm}  +10 n a_{n} c_s  \left(3 \Bbbk r_{\text{\tiny 0}}+2 c_s (26 \beta -34 i n-51 i)\right)\underbrace{\rho ^{n+3}}+4 i n a_{n} c_s \left(7 c_s (30 i \beta +31 n+62)+3 i \Bbbk r_{\text{\tiny 0}}\right)\underbrace{ \rho ^{n+4} }   \\ \nonumber   
& \hspace{0.5cm}+2 n a_{n} c_s \left(\Bbbk r_{\text{\tiny 0}}+c_s (448\beta -388 i n-970 i)\right)\underbrace{\rho ^{n+5} } +i n a_{n} c_s^2 (656 i \beta +493 n+1479) \underbrace{\rho ^{n+6}} +110 n a_{n} c_s^2 (3 \beta -2 i n-7 i) \underbrace{\rho ^{n+7}}  \\ \nonumber 
& \hspace{0.5cm} +22 i n a_{n} c_s^2 (5 i \beta +3 n+12) \underbrace{\rho ^{n+8}}  +2 n a_{n} c_s^2 (11 \beta -6 i n-27 i) \underbrace{\rho ^{n+9}}+i n a_{n} c_s^2 (2 i \beta
   +n+5) \underbrace{\rho ^{n+10}} -16 n^2 a_{n}^2 c_s^2 \underbrace{\rho ^{2 n}}+112 n^2 a_{n}^2 c_s^2 \underbrace{\rho ^{2 n+1}} \\ \nonumber 
& \hspace{0.5cm} -356 n^2 a_{n}^2 c_s^2 \underbrace{\rho ^{2 n+2}}+680 n^2 a_{n}^2 c_s^2 \underbrace{\rho^{2 n+3}}-868 n^2 a_{n}^2 c_s^2 \underbrace{\rho ^{2 n+4}}  + 776 n^2 a_{n}^2 c_s^2 \underbrace{\rho ^{2 n+5}}-493 n^2 a_{n}^2 c_s^2 \underbrace{\rho ^{2 n+6}}  +220 n^2 a_{n}^2 c_s^2 \underbrace{\rho ^{2 n+7}}  \\ \nonumber
& \hspace{0.5cm} -66 n^2 a_{n}^2 c_s^2 \underbrace{\rho ^{2 n+8}}+12 n^2 a_{n}^2 c_s^2 \underbrace{\rho ^{2 n+9}}-n^2 a_{n}^2 c_s^2 \underbrace{\rho ^{2 n+10}} \hspace{0.2cm}\Bigg]  \\ \nonumber  
& = \hspace{0.5cm}\underbrace{-\omega^2 r_{\text{\tiny 0}}^2 }+4 c_s \left(c_s (4 i \beta +\bar{l}+4)+\mathcal{M}^2 r_{\text{\tiny 0}}^2 c_s-2 \beta  \Bbbk r_{\text{\tiny 0}}\right)\underbrace{\rho}  -2 c_s \left(c_s \left(-8 \beta ^2+68 i \beta +7 \bar{l}+60\right)+3 \mathcal{M}^2 r_{\text{\tiny 0}}^2 c_s-10 \beta 
   \Bbbk r_{\text{\tiny 0}}\right)\underbrace{\rho ^2}  \\ \nonumber  
& \hspace{0.5cm} +4 c_s \left(\mathcal{M}^2 r_{\text{\tiny 0}}^2 c_s+5 \left(-\beta  \Bbbk r_{\text{\tiny 0}}+c_s \left(-4 \beta ^2+24 i \beta
   +\bar{l}+20\right)\right)\right)\underbrace{\rho ^3}  - c_s \left(\mathcal{M}^2 r_{\text{\tiny 0}}^2 c_s+5 \left(-2 \beta  \Bbbk r_{\text{\tiny 0}}+3 c_s \left(-12 \beta ^2+64 i \beta +\bar{l}+52\right)\right)\right)\underbrace{\rho ^4} \\ \nonumber  
 & \hspace{0.5cm} + 2 c_s \left(-\beta  \Bbbk r_{\text{\tiny 0}}+3 c_s \left(-40 \beta ^2+204 i \beta +\bar{l}+164\right)\right)\underbrace{\rho ^5}  + c_s^2 \left(208 \beta^2-1044 i \beta -\bar{l}-836\right)\underbrace{\rho ^6} - \left(\beta ^2-5 i \beta -4\right)\Big\lbrace 120 c_s^2  \underbrace{\rho ^7 }+45 c_s^2  \underbrace{\rho ^8 }  \\ \label{eq:A9}
& \hspace{15cm}  -10 c_s^2 \underbrace{\rho ^9 } +  c_s^2 \underbrace{\rho ^{10}} \Big\rbrace .
\end{align}
\end{footnotesize}
Each power of $\rho$, in the above equation, is again purposely under-braced to depict the whole picture as vividly as possible in front of the reader.
\par 
In the following calculation, we would go on reducing Eq.\eqref{eq:aa53} and evaluate the summations systematically. The intermediate steps are shown here to arrive at Eq.\eqref{eq:aa54} in Section 4.2 .
From Eq.\eqref{eq:aa53}, \\
\\
\begin{small}
\be \label{eq:A10}
\mathcal{S}_{1} = \sum_{\mathclap{n=1}}^{\infty}\hspace{0.1cm}\sum_{\mathclap{k=0}}^{10}\rho^{n+k} \hspace{0.1cm}\mathfrak{f}_{k}^{\mathbb{I}}(n)
 = \left[ \hspace{0.05cm}\sum_{\mathclap{n=1}}^{\infty}\rho^{n+0} \hspace{0.1cm}\mathfrak{f}_{0}^{\mathbb{I}}(n) +  \sum_{\mathclap{n=1}}^{\infty}\rho^{n+1} \hspace{0.1cm}\mathfrak{f}_{1}^{\mathbb{I}}(n) +  \sum_{\mathclap{n=1}}^{\infty}\rho^{n+2} \hspace{0.1cm}\mathfrak{f}_{2}^{\mathbb{I}}(n) + \hdots + \sum_{\mathclap{n=1}}^{\infty}\rho^{n+10} \hspace{0.1cm}\mathfrak{f}_{10}^{\mathbb{I}}(n) \right].
\ee
\end{small}
 \end{widetext}
 
For any particular $0\leqslant k\leqslant 10$, the general term being singled out from $\mathcal{S}_{1}$ is
\begin{small}
\begin{align} \nonumber
& \hspace{0.5cm}\sum_{\mathclap{n=1}}^{\infty}\rho^{n+k} \hspace{0.1cm}\mathfrak{f}_{k}^{\mathbb{I}}(n) \\ \nonumber \\ \nonumber
& = \sum_{\mathclap{n=1}}^{12-k-1}\rho^{n+k} \hspace{0.1cm}\mathfrak{f}_{k}^{\mathbb{I}}(n) \hspace{0.5cm}+ \underbrace{\sum_{\mathclap{n=12-k}}^{\infty}\rho^{n+k} \hspace{0.1cm}\mathfrak{f}_{k}^{\mathbb{I}}(n)}_{\substack{\boxed{n+k=\lambda + 12, \hspace{0.1cm}\text{say}}}}  \\ \label{eq:A11}
& = \underbrace{\sum_{\mathclap{n=1}}^{12-k-1}\rho^{n+k} \hspace{0.1cm}\mathfrak{f}_{k}^{\mathbb{I}}(n)}_{\text{finite sum} \hspace{0.1cm}\forall \hspace{0.05cm}k} \hspace{0.2cm}+\hspace{0.2cm} \underbrace{\sum_{\mathclap{\lambda=0,1,..}}^{\infty}\rho^{\lambda+12} \hspace{0.2cm} \hspace{0.1cm}\mathfrak{f}_{k}^{\mathbb{I}}(\lambda+12-k)}_{\boxed{\lambda + 12=j, \hspace{0.1cm}\text{say}}}. 
\end{align}
\be \label{eq:A12}
\therefore \hspace{0.2cm} \mathcal{S}_{1}=\sum_{\mathclap{n=1}}^{12-k-1}\sum_{\mathclap{k=0}}^{10}\rho^{n+k} \hspace{0.1cm}\mathfrak{f}_{k}^{\mathbb{I}}(n) \hspace{0.15cm}+\hspace{0.15cm} \underbrace{\sum_{\mathclap{j=12,13,..}}^{\infty}\hspace{0.1cm}\Bigg(\sum_{\mathclap{k=0}}^{10} \hspace{0.1cm}\mathfrak{f}_{k}^{\mathbb{I}}(j-k)\Bigg)\rho^{j}}_{\mathfrak{s}_{1}\hspace{0.2cm}(\text{say})}.
\ee
\end{small}
Similarly like Eq.\eqref{eq:A10},
\begin{small}
\begin{align} \nonumber
&\mathcal{S}_{2} = \sum_{\mathclap{n=1}}^{\infty}\hspace{0.1cm}\sum_{\mathclap{p=0}}^{10}\rho^{2n+p} \hspace{0.1cm}\mathfrak{f}_{p}^{\mathbb{II}}(n) \\ \nonumber \\ \nonumber
&= \Bigg[\underbrace{\sum_{\mathclap{n=1}}^{\infty}\rho^{2n+0} \hspace{0.1cm}\mathfrak{f}_{1}^{\mathbb{II}}(n)+\sum_{\mathclap{n=1}}^{\infty}\rho^{2n+2} \hspace{0.1cm}\mathfrak{f}_{2}^{\mathbb{II}}(n)+...+\sum_{\mathclap{n=1}}^{\infty}\rho^{2n+10} \hspace{0.1cm}\mathfrak{f}_{10}^{\mathbb{II}}(n)}_{\text{say},\hspace{0.2cm}\mathcal{S}_{2}^{\mathbb{I}} \hspace{0.1cm} \text{where} \hspace{0.1cm} \forall \hspace{0.05cm}p\equiv p_{1}=0,2,...,10} \\ \nonumber \\ \label{eq:A13}
& \hspace{0.5cm} +  \underbrace{\sum_{\mathclap{n=1}}^{\infty}\rho^{2n+1} \hspace{0.1cm}\mathfrak{f}_{1}^{\mathbb{II}}(n)+\sum_{\mathclap{n=1}}^{\infty}\rho^{2n+3} \hspace{0.1cm}\mathfrak{f}_{3}^{\mathbb{II}}(n)+...+\sum_{\mathclap{n=1}}^{\infty}\rho^{2n+9} \hspace{0.1cm}\mathfrak{f}_{9}^{\mathbb{II}}(n)}_{\text{say},\hspace{0.2cm}\mathcal{S}_{2}^{\mathbb{II}}\hspace{0.1cm} \text{where} \hspace{0.1cm} \forall \hspace{0.05cm}p\equiv p_{2}=1,3,...,9} \Bigg]. 
\end{align}
\end{small}
Now the general term from the above $\mathcal{S}_{2}^{\mathbb{I}}$ is singled out as the following,
\begin{small}
\begin{align} \nonumber
 & \sum_{\mathclap{n=1}}^{\infty}\rho^{2n+p_{1}} \hspace{0.1cm}\mathfrak{f}_{p_{1}}^{\mathbb{II}}(n) \\ \nonumber \\ \nonumber
 &= \sum_{\mathclap{n=1}}^{\frac{12-p_{1}}{2}-1}\rho^{2n+p_{1}} \hspace{0.1cm}\mathfrak{f}_{p_{1}}^{\mathbb{II}}(n) \hspace{0.3cm}+\hspace{0.3cm} \underbrace{\sum_{\mathclap{n=\frac{12-p_{1}}{2}}}^{\infty}\rho^{2n+p_{1}} \hspace{0.1cm}\mathfrak{f}_{p_{1}}^{\mathbb{II}}(n)}_{\substack{\boxed{2n+p_{1}=\lambda_{1} + 12, \hspace{0.1cm}\text{say}}}}   \\ \label{eq:A14}
 & = \underbrace{\sum_{\mathclap{n=1}}^{\frac{12-p_{1}}{2}-1}\rho^{2n+p_{1}} \hspace{0.1cm}\mathfrak{f}_{p_{1}}^{\mathbb{II}}(n)}_{\text{finite sum} \hspace{0.1cm}\forall \hspace{0.05cm}p_{1} } \hspace{0.15cm}+\hspace{0.15cm} \underbrace{\sum_{\mathclap{\lambda_{1}=0,2,..}}^{\infty}\rho^{\lambda_{1}+12} \hspace{0.2cm} \hspace{0.1cm}\mathfrak{f}_{p_{1}}^{\mathbb{II}}\left(\frac{\lambda_{1}+12-p_{1}}{2}\right)}_{\boxed{\lambda_{1} + 12=j, \hspace{0.1cm}\text{say}}}. 
\end{align} 
\end{small}
\begin{small}
 \begin{align} \nonumber
& \therefore \hspace{0.2cm} \mathcal{S}_{2}^{\mathbb{I}}=\sum_{\mathclap{n=1}}^{\frac{12-p_{1}}{2}-1}\sum_{\mathclap{p_{1}=0,2,..}}^{10}\rho^{2n+p_{1}} \hspace{0.1cm}\mathfrak{f}_{p_{1}}^{\mathbb{II}}(n)  \\  \label{eq:A15} 
& \hspace{3.5cm}+ \hspace{0.15cm}\underbrace{\sum_{\mathclap{j=12,14,..}}^{\infty}\hspace{0.1cm}\Bigg(\hspace{0.3cm}\sum_{\mathclap{p_{1}=0,2,..}}^{10} \hspace{0.1cm}\mathfrak{f}_{p_{1}}^{\mathbb{II}}\left(\frac{j-p_{1}}{2} \right)\hspace{0.2cm}\Bigg)\rho^{j} }_{\mathfrak{s}_{2}\hspace{0.2cm}(\text{say})}.
\end{align}
\end{small}
Again, one can single out the general term from $\mathcal{S}_{2}^{\mathbb{II}}$ as well,
\begin{small}
\begin{align} \nonumber
& \sum_{\mathclap{n=1}}^{\infty}\rho^{2n+p_{2}} \hspace{0.1cm}\mathfrak{f}_{p_{2}}^{\mathbb{II}}(n) \\ \nonumber \\ \nonumber
 & = \sum_{\mathclap{n=1}}^{\frac{13-p_{2}}{2}-1}\rho^{2n+p_{2}} \hspace{0.1cm}\mathfrak{f}_{p_{2}}^{\mathbb{II}}(n) \hspace{0.3cm}+\hspace{0.3cm} \underbrace{\sum_{\mathclap{n=\frac{13-p_{2}}{2}}}^{\infty}\rho^{2n+p_{2}} \hspace{0.1cm}\mathfrak{f}_{p_{2}}^{\mathbb{II}}(n) }_{\substack{\boxed{2n+p_{2}=\lambda_{2} + 12, \hspace{0.1cm}\text{say}}}} \\ \label{eq:A16}
 & = \underbrace{\sum_{\mathclap{n=1}}^{\frac{13-p_{2}}{2}-1}\rho^{2n+p_{2}} \hspace{0.1cm}\mathfrak{f}_{p_{2}}^{\mathbb{II}}(n)}_{\text{finite sum} \hspace{0.1cm}\forall \hspace{0.05cm}p_{2}} \hspace{0.15cm}+\hspace{0.15cm} \underbrace{\sum_{\mathclap{\lambda_{2}=1,3,..}}^{\infty}\rho^{\lambda_{2}+12} \hspace{0.2cm} \hspace{0.1cm}\mathfrak{f}_{p_{2}}^{\mathbb{II}}\left(\frac{\lambda_{2}+12-p_{2}}{2}\right)}_{\boxed{\lambda_{2} + 12=j, \hspace{0.1cm}\text{say}}}. \\ \nonumber
& \therefore \hspace{0.2cm} \mathcal{S}_{2}^{\mathbb{II}}=\sum_{\mathclap{n=1}}^{\frac{13-p_{2}}{2}-1}\sum_{\mathclap{p_{2}=1,3,..}}^{9}\rho^{2n+p_{2}} \hspace{0.1cm}\mathfrak{f}_{p_{2}}^{\mathbb{II}}(n) \\ \label{eq:A17}
& \hspace{3.5cm}+ \hspace{0.15cm} \underbrace{\sum_{\mathclap{j=13,15,..}}^{\infty}\hspace{0.1cm}\Bigg(\hspace{0.3cm}\sum_{\mathclap{p_{2}=1,3,..}}^{9} \hspace{0.1cm}\mathfrak{f}_{p_{2}}^{\mathbb{II}}\left(\frac{j-p_{2}}{2}\right)\hspace{0.2cm}\Bigg)\rho^{j} }_{\mathfrak{s}_{3}\hspace{0.2cm}(\text{say})}.
\end{align}
\end{small}

One can notice that,
\begin{itemize}
\item
for $\mathfrak{s}_{2}$ in Eq.\eqref{eq:A15}, the necessary condition is: $(j-p_{1})$ has to always be even with $\forall \hspace{0.05cm}p_{1}=0,2,...,10$ and $\forall \hspace{0.05cm}j=12,14,...,\infty$ and 
\item
for $\mathfrak{s}_{3}$ in Eq.\eqref{eq:A17}, the necessary condition is: $(j-p_{2})$ has to always be even with $\forall \hspace{0.05cm}p_{2}=1,3,...,9$ and $\forall \hspace{0.05cm}j=13,15,...,\infty$.
\end{itemize}
Clearly the $j$-s are different in $\mathfrak{s}_{2}$ and $\mathfrak{s}_{3}$, but since the corresponding $\rho^{j}$-s in Eqs.\eqref{eq:A15} and \eqref{eq:A17} are all linearly independent, we can practically club these two separate infinite summations into a single one. Thus

\begin{small}
\begin{align} \nonumber
& \mathcal{S}_{2}=\left(\mathcal{S}_{2}^{\mathbb{I}}+\mathcal{S}_{2}^{\mathbb{II}} \right) \xRightarrow [\text{summations}]{\text{infinite}} \left(\mathfrak{s}_{2} + \mathfrak{s}_{3}\right) \\ \nonumber \\ \nonumber
&= \hspace{0.2cm} \sum_{\mathclap{j=12,14,..}}^{\infty}\hspace{0.1cm}\Bigg(\hspace{0.3cm}\sum_{\mathclap{p_{1}=0,2,..}}^{10} \hspace{0.1cm}\mathfrak{f}_{p_{1}}^{\mathbb{II}}\left(\frac{j-p_{1}}{2} \right)\hspace{0.05cm}\Bigg)\rho^{j} \\ \nonumber
& \hspace{3.5cm}+ \hspace{0.15cm} \sum_{\mathclap{j=13,15,..}}^{\infty}\hspace{0.1cm}\Bigg(\hspace{0.3cm}\sum_{\mathclap{p_{2}=1,3,..}}^{9} \hspace{0.1cm}\mathfrak{f}_{p_{2}}^{\mathbb{II}}\left(\frac{j-p_{2}}{2}\right)\hspace{0.05cm}\Bigg)\rho^{j} \\ \nonumber \\ \nonumber
&= \hspace{0.3cm}\underbrace{\sum_{\mathclap{j=12,13,..}}^{\infty}\hspace{0.1cm}\Bigg(\hspace{0.2cm}\sum_{\mathclap{p=0,1,..}}^{10} \hspace{0.1cm}\mathfrak{f}_{p}^{\mathbb{II}}\left(\frac{j-p}{2} \right)\hspace{0.05cm}\Bigg)\rho^{j} }_{\mathfrak{s}_{2,3}\hspace{0.2cm}(\text{say})} \hspace{0.5cm} \big(\text{provided}\hspace{0.1cm} \forall \hspace{0.05cm}(j-p)=0,2,4,... \big) . \\ \label{eq:A18}
\end{align}
\end{small}

And through Eqs.\eqref{eq:A12} and \eqref{eq:A18}
\begin{small}
\begin{align} \nonumber
& \left(\mathcal{S}_{1}+\mathcal{S}_{2} \right) \xRightarrow [\text{summations}]{\text{infinite}} \left(\mathfrak{s}_{1} + \mathfrak{s}_{2,3}\right) \\ \nonumber \\ \nonumber
&=\hspace{0.2cm} \sum_{\mathclap{j=12,13,..}}^{\infty}\hspace{0.1cm}\Bigg(\sum_{\mathclap{k=0}}^{10} \hspace{0.1cm}\mathfrak{f}_{k}^{\mathbb{I}}(j-k)\Bigg)\rho^{j} \\ \nonumber 
& \hspace{3.5cm}+ \hspace{0.15cm} \sum_{\mathclap{j=12,13,..}}^{\infty}\hspace{0.1cm}\Bigg(\hspace{0.2cm}\sum_{\mathclap{p=0,1,..}}^{10} \hspace{0.1cm}\mathfrak{f}_{p}^{\mathbb{II}}\left(\frac{j-p}{2} \right)\hspace{0.2cm}\Bigg)\rho^{j}  \\ \nonumber \\ \nonumber
&= \hspace{0.2cm} \sum_{\mathclap{j=12,13,...}}^{\infty}\hspace{0.35cm}\left[\hspace{0.1cm}\sum_{\mathclap{k=0}}^{10} \hspace{0.1cm}\mathfrak{f}_{k}^{\mathbb{I}}(j-k) \hspace{0.3cm}+\hspace{0.3cm} \underbrace{\sum_{\mathclap{p=0,1,..}}^{10} \hspace{0.1cm}\mathfrak{f}_{p}^{\mathbb{II}}\left(\frac{j-p}{2} \right)\hspace{0.2cm}}_{\forall \hspace{0.05cm}(j-p)=0,2,4,...,\infty}\right]\rho^{j}. \\ \label{eq:A19}
\end{align}
\end{small}
After having these Eqs.\eqref{eq:A12}, \eqref{eq:A15}, \eqref{eq:A17} and \eqref{eq:A19} clubbed together, we had written the l.h.s of Eq.\eqref{eq:aa54} where the last square bracket now generates a recursion relation quite naturally. This helps find out the Frobenius coefficient(s) for any arbitrary $j=12,13,14,...$ provided $(j-p)=0,2,4,...$ holds true $\forall p=0,1,...,10$. The recursion relation is explicitly given by (please turn over this page),

\begin{widetext}
\begin{small}
\bea \nonumber
& a_{j}=  -\frac{1}{32j\left(2c_{s}j+i \omega r_{\text{\tiny 0}} \right) }
\Bigg[a_{j-1}\Big(-16(j-1)\big\lbrace 28 j c_s -14 c_s +i \left(8 \beta  c_s+7 r_{\text{\tiny 0}} \omega \right)\big\rbrace\Big)   \\ \nonumber
& + a_{j-2}\Big(16(j-2)\big\lbrace 89 j c_s -89 c_s +i \left(48 \beta  c_s+10 r_{\text{\tiny 0}} \omega \right) \big\rbrace \Big)
+ a_{j-3}\Big(-40(j-3)\big\lbrace 68 j c_s -102 c_s  +i \left(52 \beta  c_s+3 r_{\text{\tiny 0}} \omega \right)\big\rbrace \Big)  \\ \nonumber
& +  a_{j-4}\Big(16(j-4)\big\lbrace 217 j c_s -434 c_s +i \left(210 \beta  c_s+3 r_{\text{\tiny 0}} \omega \right) \big\rbrace \Big)
+ a_{j-5}\Big(-8(j-5)\big\lbrace 388 j c_s -970 c_s +i \left(448 \beta  c_s+r_{\text{\tiny 0}} \omega \right) \big\rbrace \Big)  \\ \nonumber
& +  a_{j-6}\Big(4(j-6)\hspace{0.05cm}c_{s}\hspace{0.05cm}\big\lbrace 493j-1479+656 i \beta \big\rbrace \Big)
+  a_{j-7}\Big(-440(j-7)\hspace{0.05cm}c_{s}\hspace{0.05cm}\big\lbrace 2j-7+3 i \beta \big\rbrace \Big) 
+  a_{j-8}\Big(88(j-8)\hspace{0.05cm}c_{s}\hspace{0.05cm}\big\lbrace 3j-12+5 i \beta \big\rbrace \Big)   \\ \nonumber
& + a_{j-9}\Big(-8(j-9)\hspace{0.05cm}c_{s}\hspace{0.05cm}\big\lbrace 6j-27+11 i \beta \big\rbrace \Big) 
+  a_{j-10}\Big(4(j-10)\hspace{0.05cm}c_{s}\hspace{0.05cm}\big\lbrace j-5+2 i \beta \big\rbrace \Big)  \\ \nonumber
& + \boxed{a_{\frac{j}{2}}^{2}}\Big(16i c_{s}j^{2}\Big) +  \boxed{a_{\frac{j-1}{2}}^{2}}\Big(-112i c_{s}(j-1)^{2}\Big) +  \boxed{a_{\frac{j-2}{2}}^{2}}\Big(356i c_{s}(j-2)^{2}\Big) + \boxed{a_{\frac{j-3}{2}}^{2}}\Big(-680i c_{s}(j-3)^{2}\Big) \\ \nonumber
& + \boxed{a_{\frac{j-4}{2}}^{2}}\Big(868i c_{s}(j-4)^{2}\Big) + \boxed{a_{\frac{j-5}{2}}^{2}}\Big(-776i c_{s}(j-5)^{2}\Big) + \boxed{a_{\frac{j-6}{2}}^{2}}\Big(493i c_{s}(j-6)^{2}\Big) + \boxed{a_{\frac{j-7}{2}}^{2}}\Big(-220i c_{s}(j-7)^{2}\Big)  \\ \label{eq:A20}
& + \boxed{a_{\frac{j-8}{2}}^{2}}\Big(66i c_{s}(j-8)^{2}\Big) + \boxed{a_{\frac{j-9}{2}}^{2}}\Big(-12i c_{s}(j-9)^{2}\Big) + \boxed{a_{\frac{j-10}{2}}^{2}}\Big(i c_{s}(j-10)^{2}\Big) \Bigg].
\eea
\end{small}
\end{widetext}

Its quite evident that the boxed coefficients written above do not contribute anything to $a_{j}$ only when $(j-p)$ is found to be an odd number $\forall p=0,1,...,10$.  Here, in Eq.\eqref{eq:A20}, $a_{j}$ is expressed in terms of the coefficients all of which are pre-determined and thus the recursion relation is consistent.
\par 
By equating the coefficients of $\rho^{2}$ on both sides of Eq.\eqref{eq:aa54}, we get the coefficient $a_{2}$,
\begin{scriptsize}
\begin{widetext}
\bea \nonumber
&\frac{1}{\left(2 c_s+i r_0 \omega \right)^2}\Bigg(4 c_s \bigg(2 \left(a_1-2 a_2\right) \left(2 c_s+i r_{\text{\tiny 0}} \omega
   \right){}^2 \left(r_{\text{\tiny 0}} \omega -4 i c_s\right)+\mathcal{M} ^4 r_{\text{\tiny 0}}^4
   c_s^3+\mathcal{M} ^2 r_{\text{\tiny 0}}^2 c_s \Big(-16 i r_{\text{\tiny 0}} \omega  c_s+2 (\bar{l} -10)
c_s^2+r_{\text{\tiny 0}}^2 \omega ^2 \Big)  +c_s \Big(-8 i (\bar{l} -4) r_{\text{\tiny 0}} \omega c_s  \\ \nonumber 
& \hspace{14cm} +(\bar{l} -12) \bar{l} c_s^2 -(\bar{l} +20) r_{\text{\tiny 0}}^2 \omega ^2 \Big)\bigg)\Bigg)=0, \\ \nonumber   
& \Rightarrow a_{2}=\frac{1}{4 \left(4 c_s^2+r_{\text{\tiny 0}}^2 \omega ^2\right){}^2 \left(16
   c_s^2+r_{\text{\tiny 0}}^2 \omega ^2\right)} \Bigg(2 a_1 \left(16 c_s^2+r_{\text{\tiny 0}}^2 \omega ^2\right) \left(4 c_s^2+r_{\text{\tiny 0}}^2
   \omega ^2\right)^2+r_{\text{\tiny 0}} \omega  c_s \Big(-\mathcal{M} ^2 r_{\text{\tiny 0}}^2
   \Big(\mathcal{M} ^2 \left(r_{\text{\tiny 0}}^4 \omega ^2 c_s^2-20 r_{\text{\tiny 0}}^2 c_s^4\right) +2(\bar{l} +44) r_{\text{\tiny 0}}^2 \omega ^2 c_s^2+8 (18-5 \bar{l} ) c_s^4+r_{\text{\tiny 0}}^4 \omega
   ^4\Big) \\ \nonumber
& \hspace{9cm}-(\bar{l}  (\bar{l} +72)+144) r_{\text{\tiny 0}}^2 \omega ^2 c_s^2+4 (\bar{l}  (5
   \bar{l} -28)-128) c_s^4+(\bar{l} +20) r_{\text{\tiny 0}}^4 \omega ^4\Big) \Bigg)  \\ \nonumber
& \hspace{1cm}   + \frac{i}{ \left(4 c_s^2+r_{\text{\tiny 0}}^2 \omega ^2\right){}^2 \left(16
   c_s^2+r_{\text{\tiny 0}}^2 \omega ^2\right)} \Bigg(2 c_s^2 \Big(\mathcal{M} ^4 \left(2 r_{\text{\tiny 0}}^4 c_s^4-r_{\text{\tiny 0}}^6 \omega ^2
   c_s^2\right)+\mathcal{M} ^2 r_{\text{\tiny 0}}^2 \left(-2 (\bar{l} +9) r_{\text{\tiny 0}}^2 \omega ^2
   c_s^2+4 (\bar{l} -10) c_s^4+r_{\text{\tiny 0}}^4 \omega ^4\right) -(\bar{l}  (\bar{l}
   +10)-40) r_{\text{\tiny 0}}^2 \omega ^2 c_s^2 \hspace{2cm}\\ \nonumber 
 & \hspace{13cm} +2 (\bar{l} -12) \bar{l}  c_s^4 +2 (\bar{l} +8)
   r_{\text{\tiny 0}}^4 \omega ^4\Big) \Bigg). \\ \label{eq:A21}
\eea
\bea \nonumber
& \therefore \hspace{0.5cm} b_{2} \equiv   a_{2} \vert_{\mathcal{M}=0}  =\Bigg(\frac{b_1}{2}+\frac{4 (\bar{l}  (5 \bar{l} -28)-128) r_{\text{\tiny 0}} \omega _1
   c_s^5-(\bar{l}  (\bar{l} +72)+144) r_{\text{\tiny 0}}^3 \omega _1^3 c_s^3+(\bar{l} +20)
   r_{\text{\tiny 0}}^5 \omega _1^5 c_s}{4 \left(4 c_s^2+r_{\text{\tiny 0}}^2 \omega
   _1^2\right)^2 \left(16 c_s^2+r_{\text{\tiny 0}}^2 \omega _1^2\right)}\Bigg)  
 +i\Bigg(\frac{-2 (\bar{l}  (\bar{l} +10)-40) r_{\text{\tiny 0}}^2 \omega _1^2 c_s^4+4 (\bar{l} +8)
   r_{\text{\tiny 0}}^4 \omega _1^4 c_s^2+4 (\bar{l} -12) \bar{l}  c_s^6}{\left(4
   c_s^2+r_{\text{\tiny 0}}^2 \omega _1^2\right)^2 \left(16 c_s^2+r_{\text{\tiny 0}}^2 \omega
   _1^2\right)}\Bigg). \\ \label{eq:A22}
\eea
\end{widetext}
\end{scriptsize}


\end{document}